\documentclass[moor]{informs1}

\usepackage[english]{babel}
\usepackage{amsmath}
\usepackage{amstext}
\usepackage{amssymb}
\usepackage{mathtools}
\usepackage{natbib}
 \NatBibNumeric
 \def\bibfont{\small}%
 \def\newblock{\ }%
 \bibpunct[, ]{[}{]}{,}{n}{}{,}%

\usepackage[colorlinks=true,breaklinks=true,bookmarks=true,urlcolor=blue,
     citecolor=blue,linkcolor=blue,bookmarksopen=false,draft=false]{hyperref}

\def\EMAIL#1{\href{mailto:#1}{#1}}

\TheoremsNumberedBySection  

\EquationsNumberedBySection 

\def\R{\mathbb{R}}

\def\rP{\mathbb{P}}

\def\Gr{\mathop{\rm Gr}}

\def\Pr{\mathop{\rm Pr}}

\def\tJ{\tilde{J}}

\def\B{{\mathcal B}}

\def\C{{\mathcal C}}

\def\P{{\mathcal P}}

\def\M{{\mathcal M}}

\def\bmu{{\boldsymbol \mu}}
\def\bnu{{\boldsymbol \nu}}
\def\bpi{{\boldsymbol \pi}}
\def\bxi{{\boldsymbol \xi}}
\def\by{{\bf y}}
\def\ba{{\bf a}}
\def\bx{{\bf x}}

\def\bu{{\bf u}}
\def\bv{{\bf v}}
\def\bs{{\bf s}}

\def\ts{{\tilde s}}
\def\tpi{{\tilde \pi}}

\def\hs{{\hat s}}

\def\sPr{{\mathsf{Pr}}}

\def\sX{{\mathsf X}}
\def\sY{{\mathsf Y}}
\def\sA{{\mathsf A}}
\def\sH{{\mathsf H}}
\def\sZ{{\mathsf Z}}
\def\sM{{\mathsf M}}
\def\sE{{\mathsf E}}

\def\sG{{\mathsf G}}
\def\sW{{\mathsf W}}
\def\sS{{\mathsf S}}
\def\sB{{\mathsf B}}
\def\sV{{\mathsf V}}
\def\sK{{\mathsf K}}

\def\tJ{{\tilde J}}

\newcommand{\appsec}{
\renewcommand{\thesubsection}{\Alph{subsection}}
}

\begin{document}

\RUNAUTHOR{Saldi, Ba\c{s}ar, and Raginsky}
\RUNTITLE{Partially Observed Stochastic Games with Mean-Field Interactions}

\TITLE{Approximate Nash Equilibria in Partially Observed Stochastic Games with Mean-Field Interactions}

\ARTICLEAUTHORS{
\AUTHOR{Naci Saldi, Tamer Ba\c{s}ar, and Maxim Raginsky}
\AFF{Coordinated Science Laboratory, University of Illinois,Urbana, IL 61801-2307, USA.\\
\{\EMAIL{nsaldi,basar1,maxim@illinois.edu}\}}}

\ARTICLEAUTHORS{
\AUTHOR{Naci Saldi}
\AFF{Department of Natural and Mathematical Sciences, Ozyegin University, Cekmekoy, Istanbul, Turkey.\\
\{\EMAIL{naci.saldi@ozyegin.edu.tr}\}}
\AUTHOR{Tamer Ba\c{s}ar, and Maxim Raginsky}
\AFF{Coordinated Science Laboratory, University of Illinois,Urbana, IL 61801-2307, USA.\\
\{\EMAIL{basar1,maxim@illinois.edu}\}}}

\ABSTRACT{
Establishing the existence of Nash equilibria for partially observed stochastic dynamic games is known to be quite challenging, with the difficulties stemming from the noisy nature of the measurements available to individual players (agents) and the decentralized nature of this information. When the number of players is sufficiently large and the interactions among agents is of the mean-field type, one way to overcome this challenge is to investigate the infinite-population limit of the problem, which leads to a mean-field game. In this paper, we consider discrete-time partially observed mean-field games with infinite-horizon discounted cost criteria. Using
the technique of converting the original partially observed stochastic control problem to a fully observed one on the belief space and the dynamic programming principle, we establish the existence of Nash equilibria for these game models under very mild technical conditions. Then, we show that the mean-field equilibrium policy, when adopted by each agent, forms an approximate Nash equilibrium for games with sufficiently many agents.
}

\MSCCLASS{91A15, 91A13, 90C40, 90C39, 60J05}
\ORMSCLASS{Primary: Games/group decisions, dynamic programming/optimal control, probability ; secondary: Stochastic, Markov, Markov processes}

\maketitle

\section{Introduction}\label{sec1}

In this paper, we consider discrete-time mean-field games with decentralized partial observation under infinite-horizon discounted-cost optimality criteria. This type of game models arise as the infinite population limit of finite-agent dynamic games of the mean-field type; that is, the interactions among agents are modeled by the mean-field term (i.e., the empirical distribution of their states), which affects both the agents' individual costs, and their state and observation transition probabilities. Letting the number of agents go to infinity, the mean-field term converges to the distribution of a single generic agent. Hence, in the limiting case, a generic agent is faced with a single-agent stochastic control problem with a constraint on the distribution of the state at each time (i.e., mean-field game problem). The main goal in this class of problems is to establish the existence of a policy and a state distribution flow such that when the generic agent applies this policy, the resulting distribution of agent's state is same as the state distribution flow. This last property is called the Nash certainty equivalence (NCE) principle (\citet{HuMaCa06}). The purpose of this paper is to study the existence of such an equilibrium for a general class of mean-field game models with discounted-cost criteria under decentralized partial observation and to establish that the policy in the mean-field equilibrium constitutes a nearly Nash equilibrium for finite-agent games with sufficiently many agents.

Mean-field games have been introduced by \citet{HuMaCa06} and \citet{LaLi07} around the same time to establish the existence of approximate Nash equilibria for fully-observed non-cooperative differential games with a large number of identical agents. The main feature of this approach is to reduce the decentralized game problem to a centralized stochastic decision problem using the NCE principle. The equilibrium solution of this decision problem provides an almost Nash equilibrium when the number of agents is sufficiently large. Characterization of the solution entails a Fokker-Planck equation evolving forward in time and a Hamilton-Jacobi-Bellman equation evolving backward in time. We refer the reader to \citet{HuCaMa07,TeZhBa14,Hua10,BeFrPh13,Ca11,CaDe13,GoSa14,MoBa16} for studies of fully-observed continuous-time mean-field games with different models and cost functions, such as games with major-minor players, risk-sensitive games, games with Markov jump parameters, and LQG games.

In the literature relatively few results are available on \emph{partially-observed} mean-field games. Indeed, our work appears to be the first one that studies discrete-time mean-field games under partial observations. Existing works have mostly studied the continuous-time setup, and analyses of continuous-time and discrete-time setups are quite different. \citet{HuCaMa06} study a partially-observed continuous-time mean-field game with linear individual dynamics. \citet{SeCa14,SeCa15} consider a continuous-time mean-field game with major-minor agents and nonlinear dynamics where the minor agents can partially observe the state of the major agent. \citet{SeCa16,SeCa16-2} also develop a nonlinear filtering theory for McKean-Vlasov type stochastic differential equations that arise as the infinite population limit of the partially-observed differential game of the mean-field type. The nonlinear filtering equation is derived using It\^{o}'s lemma for Banach space valued stochastic processes. \citet{TaMe16} study a continuous-time partially observed stochastic control problem of the mean-field type and establish a maximum principle to characterize the optimal control. \citet{HuWa14} consider a continuous-time mean-field game with linear individual dynamics where two types of partial information structure are considered: (i) agents cannot observe the white-noise which is common to all agents, (ii) agents can access the additive white-noise version of their own states.

The class of discrete-time mean-field games we consider in this paper are defined on a Polish state space and with infinite-horizon discounted-cost optimality criteria for the players, who have access to decentralized partial observation on their individual states. In such games, a generic agent is faced with a partially observed stochastic control problem under the NCE principle, which, as indicated earlier, states that the state distribution flow under an optimal decision rule should be the same as the mean-field term that appears in the state and observation dynamics as well as in the individual cost functions. In accordance with this, the classical techniques used to study partially observed optimal control problems are not sufficient to analyze mean-field games. To establish the existence of an equilibrium solution, we have to bring in the fixed-point approach that is used to obtain equilibria in classical game problems, along with the technique of converting partially observed optimal control problems to fully observed ones on the belief space and then employing the dynamic programming principle. The precise descriptions of the mean-field game and the finite-agent game problems are given in Sections~\ref{sec3} and \ref{sec2}, respectively. In Section~\ref{main-proof} we prove the existence of a mean-field equilibrium. In Section~\ref{sec4} we establish that the mean-field equilibrium policy is approximately Nash for finite-agent games with sufficiently many agents. In Section~\ref{example} we illustrate our results by considering an example.

In \citet{SaBaRa17} (see also the abridged conference version \cite{SaBaRa17-ACC}) we solved the fully-observed version of this problem under a similar set of assumptions on the system components. The techniques used in this paper to establish the existence of a mean field and an approximate Nash equilibrium are almost the same as in \citet{SaBaRa17} modulo some transformations of the original problems into equivalent ones for which we can still use the techniques in \citet{SaBaRa17}. However, as a result of these transformations, there are highly non-trivial differences between the proofs in this paper and in \citet{SaBaRa17}. For instance, as a result of fully observed reduction of partially observed optimal control problem in the mean field game, the dependence of the state transition probability of the fully observed model on the mean-field term is not explicit as in \citet{SaBaRa17}. Therefore, it is quite challenging to prove the weak continuity of this transition probability
with respect to the mean-field term, which is a very crucial result in order to establish the existence of a mean-field equilibrium.


\smallskip

\noindent\textbf{Notation.} For a metric space $\sE$, we let $C_b(\sE)$ denote the set of all bounded continuous real functions on $\sE$, and $\P(\sE)$ denote the set of all Borel probability measures on $\sE$. For any $\sE$-valued random element $x$, ${\cal L}(x)(\,\cdot\,) \in \P(\sE)$ denotes the distribution of $x$. A sequence $\{\mu_n\}$ of measures on $\sE$ is said to converge weakly to a measure $\mu$ if $\int_{\sE} g(e) \mu_n(de)\rightarrow\int_{\sE} g(e) \mu(de)$ for all $g \in C_b(\sE)$. For any $\nu \in \P(\sE)$ and measurable real function $g$ on $\sE$, we define $\nu(g) \coloneqq \int g d\nu$. For any subset $B$ of $\sE$, we let $\partial B$ and $B^c$ denote the boundary and complement of $B$, respectively. The notation $v\sim \nu$ means that the random element $v$ has distribution $\nu$. Unless otherwise specified, the term ``measurable" will refer to Borel measurability.

\section{Finite Player Game}\label{sec2}


We consider a discrete-time partially observed $N$-agent stochastic game of mean-field type with a Polish state space $\sX$, a Polish action space $\sA$, and a Polish observation space $\sY$. For every $t \in \{0,1,2,\ldots\}$ and every $i \in \{1,2,\ldots,N\}$, let $x^N_i(t) \in \sX$, $a^N_i(t) \in \sA$, and $y^N_i(t) \in \sY$ denote the state, the action, and the observation of Agent~$i$ at time $t$, and let
\begin{align}
e_t^{(N)}(\,\cdot\,) \coloneqq \frac{1}{N} \sum_{i=1}^N \delta_{x_i^N(t)}(\,\cdot\,) \in \P(\sX) \nonumber
\end{align}
denote the empirical distribution of the states of agents at time $t$, where $\delta_x\in\P(\sX)$ is the Dirac measure at $x$. The initial states $x^N_i(0)$ are independent and identically distributed according to $\mu_0$. For each $t \ge 0$, the current-observations $(y^N_1(t),\ldots,y^N_N(t))$ and the next-states  $(x^N_1(t+1),\ldots,x^N_N(t+1))$ of agents are obtained randomly according to the conditional probability laws
\begin{align}
&\prod^N_{i=1} r\big(dy^N_i(t)\big|x^N_i(t),e^{(N)}_t\big),  \nonumber \\ 
&\prod^N_{i=1} p\big(dx^N_i(t+1)\big|x^N_i(t),a^N_i(t),e^{(N)}_t\big), \nonumber 
\end{align}
where the stochastic kernel $p : \sX \times \sA \times \P(\sX) \to \P(\sX)$ denotes the transition probability law of the next state given the previous state-action pair and the empirical distribution of states, and the stochastic kernel $r: \sX \times \P(\sX) \to \P(\sY)$ denotes the transition probability law of the current observation given the current state and the empirical distribution of states. The measurable function $c: \sX \times \sA \times \P(\sX) \rightarrow [0,\infty)$ is the one-stage cost function.

Define the history spaces $\sH_0 = \sY$ and $\sH_{t}= (\sY\times\sA)^t\times\sY$ for $t=1,2,\ldots$, all endowed with product Borel $\sigma$-algebras. A \emph{policy} for a generic agent is a sequence $\pi=\{\pi_{t}\}$ of stochastic kernels on $\sA$ given $\sH_{t}$. The set of all policies for Agent~$i$ is denoted by $\Pi_i$. Let $\tilde{\Pi}_i$ be the set of policies in $\Pi_i$ which only use the observations; that is, $\pi \in \tilde{\Pi}_i$ if $\pi_t: \prod_{k=0}^t \sY \rightarrow \P(\sA)$ for each $t\geq0$. Let ${\bf \Pi}^{(N)} = \prod_{i=1}^N \Pi_i$ and $\tilde{{\bf \Pi}}^{(N)} = \prod_{i=1}^N \tilde{\Pi}_i$. We let ${\boldsymbol \pi}^{(N)} \coloneqq (\pi^1,\ldots,\pi^N)$, $\pi^i \in \Pi_i$ denote the $N$-tuple of policies for all the agents in the game. Under such $N$-tuple of policies, the actions of agents at each time $t \ge 0$ is obtained randomly according to the conditional probability law
\begin{align}\label{eq:policy_spec}
\prod^N_{i=1} \pi^i_t\big(da^N_i(t)\big|h^N_i(t)\big),
\end{align}
where $h^N_i(t) = (y^N_i(t),a^N_i(t-1),y^N_i(t-1)\ldots,a^N_i(0),y^N_i(0))$ is the observation-action history observed by Agent~$i$ up to time $t$. Note that agents can only use their local observations. Hence, it is a partially observed game model.

For Agent~$i$, the infinite-horizon discounted cost under the initial distribution $\mu_0$ and $N$-tuple of policies ${\boldsymbol \pi}^{(N)} \in {\bf \Pi}^{(N)}$ is given by
\begin{align}
J_i^{(N)}({\boldsymbol \pi}^{(N)}) &= E^{{\boldsymbol \pi}^{(N)}}\biggl[\sum_{t=0}^{\infty}\beta^{t}c(x_{i}^N(t),a_{i}^N(t),e^{(N)}_t)\biggr], \nonumber
\end{align}
where $E^{{\boldsymbol \pi}^{(N)}}\big[\cdot\big]$ denotes expectation with respect to the unique probability law induced by the above stochastic update rules and the initial state distribution $\mu_0$ on the infinite product of state, observation, and action spaces of all agents.

\begin{definition}
A policy ${\boldsymbol \pi}^{(N*)}= (\pi^{1*},\ldots,\pi^{N*})$ constitutes a \emph{Nash equilibrium} if
\begin{align}
J_i^{(N)}({\boldsymbol \pi}^{(N*)}) = \inf_{\pi^i \in \Pi_i} J_i^{(N)}({\boldsymbol \pi}^{(N*)}_{-i},\pi^i) \nonumber
\end{align}
for each $i=1,\ldots,N$, where ${\boldsymbol \pi}^{(N*)}_{-i} \coloneqq (\pi^{j*})_{j\neq i}$.
\end{definition}

Establishing the existence of Nash equilibria for the class of games formulated above is known to be
difficult due to (almost) decentralized and noisy nature of the information structure of the problem. Indeed, even if the number of players is small, it is all but impossible to show even the existence of approximate Nash equilibria for these types of games. However, when the number of players is sufficiently large, one way to overcome this challenge is to introduce the infinite-population limit $N\rightarrow\infty$ of the game described here. In this limiting case, we can model the empirical distributions of the state configurations as an exogenous state-measure flow, which should be consistent with the distribution of a generic agent (i.e., the NCE principle) by the law of large numbers. Hence, in the limiting case, a generic agent is faced with a mean-field game that will be introduced in the next section. Then one would expect that if each agent in the finite-agent $N$ game problem adopts the equilibrium policy in the infinite-population limit, the resulting policy will be an approximate Nash equilibrium for all sufficiently large $N$. Therefore, by studying the infinite-population limit, which is easier to handle, one can obtain approximate Nash equilibrium for the original finite-agent game problem for which establishing the existence of a true Nash equilibrium is very difficult.

To that end, we slightly change the definition of Nash equilibrium in this model and adopt the following solution concept:
\begin{definition}\label{def1}
A policy ${\boldsymbol \pi}^{(N*)} \in \tilde{{\bf \Pi}}^{(N)}$ is a \emph{Nash equilibrium} if
\begin{align*}
J_i^{(N)}({\boldsymbol \pi}^{(N*)}) &= \inf_{\pi^i \in \tilde{\Pi}_i} J_i^{(N)}({\boldsymbol \pi}^{(N*)}_{-i},\pi^i)
\end{align*}
for each $i=1,\ldots,N$, and an \emph{$\varepsilon$-Nash equilibrium} (for a given $\varepsilon > 0$) if
\begin{align*}
J_i^{(N)}({\boldsymbol \pi}^{(N*)}) &\leq \inf_{\pi^i \in \tilde{\Pi}_i} J_i^{(N)}({\boldsymbol \pi}^{(N*)}_{-i},\pi^i) + \varepsilon
\end{align*}
for each $i=1,\ldots,N$.
\end{definition}

Note that, according to this definition, the agents can only use their local observations $(y^N_i(t),\ldots,y^N_i(0))$ to design their policies. Indeed, in practical applications, agents typically have access only to their local observations. Collecting all the observations in the entire system is intractable, particularly if the number of agents is large. Consequently, establishing that a mean-field policy is an approximate Nash equilibrium for the game under the assumption that the agents have access to full observation variables is not necessary in order to cover practically meaningful scenarios. It is sufficient to establish the existence of an approximate Nash equilibrium for the game with a local information structure. In addition, in the discrete-time mean field literature, it is common to establish the existence of approximate Nash equilibria with local (decentralized) information structures (see \citet{AdJoWe15} \citet{Bis15}). This is true for continuous-time partially observed case as well (see \citet{SeCa16-3}).

In Section~\ref{sec4}, we will show that the policy in the infinite-population equilibrium is an $\varepsilon$-Nash equilibrium (for a given $\varepsilon > 0$) when the number of agents is sufficiently large.

\section{Partially observed mean-field games and mean-field equilibria}\label{sec3}

In this section we introduce a mean-field game that can be interpreted as the infinite-population limit $N\rightarrow\infty$ of the game introduced in the preceding section. This mean-field game model is specified by
\begin{align}
\bigl( \sX, \sA, \sY, p, r, c, \mu_0 \bigr), \nonumber
\end{align}
where, as before, $\sX$, $\sY$, and $\sA$ are the Polish state, observation, and action spaces, respectively. The stochastic kernel $p : \sX \times \sA \times \P(\sX) \to \P(\sX)$ denotes the transition probability and $r: \sX \times \P(\sX) \to \P(\sY)$ denotes the observation kernel. The measurable function $c: \sX \times \sA \times \P(\sX) \rightarrow [0,\infty)$ is the one-stage cost function and $\mu_0$ is the distribution of the initial state.

Define the history spaces $\sG_0 = \sY$ and $\sG_{t}=(\sY\times\sA)^{t}\times \sY$ for $t=1,2,\ldots$, all endowed with product Borel $\sigma$-algebras. A \emph{policy} is a sequence $\pi=\{\pi_{t}\}$ of stochastic kernels on $\sA$ given $\sG_{t}$. The set of all policies is denoted by $\Pi$. Partially observed mean-field games are not games in the classical sense. They are indeed single-agent partially observed stochastic control problems whose state distribution at each time step should satisfy some consistency condition. In other words, we have a single agent with partial observations and model the overall behavior of (a large population of) other agents by an exogenous \textit{state-measure flow} $\bmu := (\mu_t)_{t \ge 0} \subset \P(\sX)$  with a given initial condition $\mu_0$. This measure flow $\bmu$ should also be consistent with the state distributions of this single agent when the agent acts optimally. The precise mathematical description of the problem is given as follows.

We let $\M \coloneqq \bigl\{\bmu \in \P(\sX)^{\infty}: \mu_0 \text{ is fixed}\bigr\}$ be the set of all state-measure flows with a given initial condition $\mu_0$. Given any measure flow $\bmu \in \M$, the probabilistic evolution of the states, observations, and actions is as follows
\begin{align}
x(0) &\sim \mu_0, \nonumber \\
y(t) &\sim p(\,\cdot\,|x(t),\mu_{t}), \text{ } t=0,1,\ldots \nonumber \\
x(t) &\sim p(\,\cdot\,|x(t-1),a(t-1),\mu_{t-1}), \text{ } t=1,2,\ldots \nonumber \\
a(t) &\sim \pi_t(\,\cdot\,|g(t)), \text{ } t=0,1,\ldots, \nonumber
\end{align}
where $g(t) \in \sG_t$ is the observation-action history  up to time $t$. According to the Ionescu Tulcea theorem (see, e.g., \citet{HeLa96}), an initial distribution $\mu_0$ on $\sX$, a policy $\pi$, and a state-measure flow $\bmu$ define a unique probability measure $P^{\pi}$ on $(\sX \times \sY \times \sA)^{\infty}$. The expectation with respect to $P^{\pi}$ is denoted by $E^{\pi}$. A policy $\pi^{*} \in \Pi$ is said to be optimal for $\bmu$ if
\begin{align}
J_{\bmu}(\pi^{*}) = \inf_{\pi \in \Pi} J_{\bmu}(\pi), \nonumber
\end{align}
where the infinite-horizon discounted cost of policy $\pi$ with measure flow $\bmu$ and the discount factor $\beta \in (0,1)$ is given by
\begin{align}
J_{\bmu}(\pi) &\coloneqq  E^{\pi}\biggl[ \sum_{t=0}^{\infty} \beta^t c(x(t),a(t),\mu_t) \biggr] \nonumber
\end{align}

Using these definitions, we first define the set-valued mapping
\begin{align}
\Psi : \M \rightarrow 2^{\Pi} \nonumber
\end{align}
as $\Psi({\boldsymbol \mu}) = \{\pi \in \Pi: \pi \text{ is optimal for } {\boldsymbol \mu}\}$.
Conversely, we define a single-valued mapping
\begin{align}
\Lambda : \Pi \to \M \nonumber
\end{align}
as follows: given $\pi \in \Pi$, the state-measure flow $\bmu := \Lambda(\pi)$ is constructed recursively as
\begin{align}
\mu_{t+1}(\,\cdot\,) = \int_{\sX \times \sA} p(\,\cdot\,|x(t),a(t),\mu_t) P^{\pi}(da(t)|x(t)) \mu_t(dx(t)), \nonumber
\end{align}
where $P^{\pi}(da(t)|x(t))$ denotes the conditional distribution of $a(t)$ given $x(t)$ under $\pi$ and $(\mu_{\tau})_{0\leq\tau\leq t}$. Using $\Psi$ and $\Lambda$, we now introduce the notion of an equilibrium for the mean-field game.

\begin{definition}
A pair $(\pi,{\boldsymbol \mu}) \in \Pi \times \M$ is a \emph{mean-field equilibrium} if $\pi \in \Psi({\boldsymbol \mu})$ and $\bmu = \Lambda(\pi)$. In other words, the policy $\pi$ is optimal for the state-measure flow $\bmu$ and $\bmu$
is consistent with the state distributions of the agent when it acts optimally via $\pi$.
\end{definition}

In this section, the main goal is to establish the existence of a mean-field equilibrium. To that end, we impose the assumptions below on the components of the mean-field game model.

\begin{assumption}\label{as1}
\begin{itemize}
\item [ ]
\item [(a)] The cost function $c$ is bounded and continuous.
\item [(b)] The stochastic kernel $p$ is weakly continuous in $(x,a,\mu)$; i.e., for all $x$, $a$ and $\mu$, $p(\,\cdot\,|x_k,a_k,\mu_k) \rightarrow p(\,\cdot\,|x,a,\mu)$ weakly when $(x_k,a_k,\mu_k) \rightarrow (x,a,\mu)$.
\item [(c)] The observation kernel $r$ is continuous in $(x,\mu)$ with respect to total variation norm; i.e., for all $x$ and $\mu$, $r(\,\cdot\,|x_k,\mu_k) \rightarrow r(\,\cdot\,|x,\mu)$ in total variation norm when $(x_k,\mu_k) \rightarrow (x,\mu)$.
\item [(d)] $\sA$ is compact and $\sX$ is locally compact.
\item [(e)] There exist a constant $\alpha \ge 0$  and a continuous moment function $w: \sX \rightarrow [0,\infty)$  (see \citet[Definition E.7]{HeLa96}) such that
\begin{align}
\sup_{(a,\mu) \in \sA \times \P(\sX)} \int_{\sX} w(y) p(dy|x,a,\mu) \leq \alpha w(x).
\end{align}
\item [(f)] The initial probability measure $\mu_0$ satisfies
\begin{align}
\int_{\sX} w(x) \mu_0(dx) \eqqcolon M < \infty. \nonumber
\end{align}
\end{itemize}
\end{assumption}

\begin{remark}
We need Assumption~\ref{as1}-(c) in order to establish the continuity of the transition probability (i.e, $\eta^{\bnu}_t(\,\cdot\,|z,a)$), that will be introduced in the next section, of the fully-observed reduction of the partially observed mean-field game model with respect to the weak topology. If this continuity condition holds under any other assumption on the observation kernel $r$ (for instance, under fully observed case; that is, $r(\,\cdot\,|x,\mu) = \delta_{x,\mu}(\,\cdot\,)$), then all the results in this paper are still true.\Halmos
\end{remark}

\noindent The main result of this section is the existence of a mean-field equilibrium under Assumption~\ref{as1}. Later we will show that this mean-field equilibrium constitutes an approximate Nash-equilibrium for games with sufficiently many agents.

\begin{theorem}\label{thm:MFE} Under Assumption~1, the mean-field game $\bigl( \sX, \sA, \sY, p, r, c, \mu_0 \bigr)$ admits a mean-field equilibrium $(\pi,\bmu)$.
\end{theorem}

The proof of Theorem~\ref{thm:MFE} is given Section~\ref{main-proof}. To establish the existence of an equilibrium, we use fully observed reduction of partially observed optimal control problems and the dynamic programming principle in addition to the fixed point approach that is commonly used in classical game problems.

\section{Proof of Theorem~\ref{thm:MFE}}\label{main-proof}

Note that, given any measure flow $\bmu \in \M$, the optimal control problem for the mean-field game reduces to one of finding an optimal policy for a partially observed Markov decision process (POMDP). Hence, before starting the proof of Theorem~\ref{thm:MFE}, we first review a few relevant results on POMDPs. To this end, fix any $\bmu \in \M$ and consider the corresponding optimal control problem.

Let $\P_w(\sX) \coloneqq \bigl\{\mu \in \P(\sX): \mu(w) < \infty \bigr\}$. It is known that any POMDP can be reduced to a (completely observable) MDP (see \citet{Yus76}, \citet{Rhe74}), whose states are the posterior state distributions or beliefs of the observer; that is, the state at time $t$ is
\begin{align}
z(t) \coloneqq \sPr\{x(t) \in \,\cdot\, | y(0),\ldots,y(t), a(0), \ldots, a(t-1)\} \in \P(\sX). \nonumber
\end{align}
We call this equivalent MDP the belief-state MDP. Note that since ${\cal L}(x(t)) \in \P_w(\sX)$ under any policy by Assumption~1-(e),(f), we have $\sPr\{x(t) \in \,\cdot\, | y(0),\ldots,y(t), a(0), \ldots, a(t-1)\} \in \P_w(\sX)$ almost everywhere. Therefore, the belief-state MDP has state space $\sZ = \P_w(\sX)$ and action space $\sA$, where $\sZ$ is equipped with the Borel $\sigma$-algebra generated by the topology of weak convergence. The transition probabilities $\{\eta_t\}_{t\geq0}$ of the belief-state MDP can be constructed as follows (see also \citet{Her89}). Let $z$ denote the generic state variable for the belief-state MDP. Fix any $t\geq 0$. First consider the transition probability on $\sX \times \sY$ given $\sZ \times \sA$
\begin{align}
R_t(x \in A, y \in B|z,a) \coloneqq \int_{\sX} \kappa_t(A,B|x',a) z(dx'), \nonumber
\end{align}
where $\kappa_t(dx,dy|x',a) \coloneqq r(dy|x,\mu_{t+1}) \otimes p(dx|x',a,\mu_t)$. Let us disintegrate $R_t$ as follows
\begin{align}
R_t(dx,dy|z,a) = H_t(dy|z,a) \otimes F_t(dx|z,a,y). \nonumber
\end{align}
Then, we define the mapping $F_t: \sZ \times \sA \times \sY \rightarrow \sZ$ as
\begin{align}
F_t(z,a,y)(\,\cdot\,) = F_t(\,\cdot\,|z,a,y) \label{eq:non_filtering}.
\end{align}
Note that
\begin{align}
F_t(z,a,y)(\,\cdot\,) &= \Pr\{x(t+1) \in \,\cdot\, | z(t) = z, a(t) = a, y(t+1) = y\} \nonumber
\intertext{and}
H_t(\,\cdot\, | z,a) &= \Pr\{y(t+1) \in \,\cdot\, | z(t) = z, a(t) = a\}. \nonumber
\end{align}
Then, $\eta_t: \sZ \times \sA \rightarrow \P(\sZ)$ can be written as
\begin{align}
\eta_t(\,\cdot\,|z(t),a(t)) = \int_{\sY} \delta_{F_t(z(t),a(t),y(t+1))}(\,\cdot\,) \text{ } H_t(dy(t+1)|z(t),a(t)). \nonumber
\end{align}
The initial point for the belief-state MDP is $\mu_0$; that is, ${\cal L}(z(0)) \sim \delta_{\mu_0}$. Finally, for each $t\geq0$, the one-stage cost function $C_t$ of the belief-state MDP is given by
\begin{align}
C_t(z,a) \coloneqq \int_{\sX} c(x,a,\mu_t) z(dx). \label{eq8}
\end{align}
Hence, the belief-state MDP is a Markov decision process with the components
\begin{align}
\bigl(\sZ,\sA,\{\eta_t\}_{t\geq0},\{C_t\}_{t\geq0},\delta_{\mu_0}\bigr). \nonumber
\end{align}

For the belief-state MDP define the history spaces $\sK_0 = \sZ$ and $\sK_{t}=(\sZ\times\sA)^{t}\times\sZ$, $t=1,2,\ldots$. A \emph{policy} is a sequence $\varphi=\{\varphi_{t}\}$ of stochastic kernels on $\sA$ given $\sK_{t}$. The set of all policies is denoted by $\Phi$. A \emph{Markov} policy is a sequence $\varphi=\{\varphi_{t}\}$ of stochastic kernels on $\sA$ given $\sZ$. The set of Markov policies is denoted by $\sM$. Let $\tJ(\varphi,\mu_0)$ denote the discounted cost function of policy $\varphi \in \Phi$ for initial point $\mu_0$ of the belief-state MDP. Notice that any history vector $s(t) = (z(0),\ldots,z(t),a(0),\ldots,a(t-1))$ of the belief-state MDP is a function of the history vector $g(t) = (y(0),\ldots,y(t),a(0),\ldots,a(t-1))$ of the POMDP. Let us write this relation as $i(g(t)) = s(t)$. Hence, for a policy $\varphi = \{\varphi_t\} \in \Phi$, we can define a policy $\pi^{\varphi} = \{\pi_t^{\varphi}\} \in \Pi$ as
\begin{align}
\pi_t^{\varphi}(\,\cdot\,|g(t)) \coloneqq \varphi_t(\,\cdot\,|i(g(t))). \nonumber
\end{align}
Let us write this as a mapping from $\Phi$ to $\Pi$: $\Phi \ni \varphi \mapsto i(\varphi) = \pi^{\varphi} \in \Pi$. It is straightforward to show that the cost functions $\tJ(\varphi,\mu_0)$ and $J(\pi^{\varphi},\mu_0)$ are the same. One can also prove that (see \citet{Yus76}, \citet{Rhe74})
\begin{align}
\inf_{\varphi \in \Phi} \tJ(\varphi,\mu_0) &= \inf_{\pi \in \Pi} J(\pi,\mu_0) \label{eq7}
\end{align}
and furthermore, that if $\varphi$ is an optimal policy for belief-state MDP, then $\pi^{\varphi}$ is optimal for the POMDP as well. Therefore, the optimal control problem for the mean-field game is equivalent to the optimal control of belief-state MDP.

Now, we derive the conditions satisfied by the components of the belief-state MDP under Assumption~1. Note first that $\sZ = \bigcup_{n\geq1} K_n$ where $K_n \coloneqq \{\mu \in \P_w(\sX): \mu(w) \leq n\}$. Since $w$ is a moment function, each $K_n$ is tight (\citet[Proposition E.8]{HeLa96}). Moreover, each $K_n$ is also closed since $w$ is continuous. Therefore, each $K_n$ is compact with respect to the weak topology. This implies that $\sZ$ is a $\sigma$-compact Polish space. Define $W:\sZ \rightarrow \R$ as
\begin{align}
W(z) = \int_{\sX} w(x) z(dx). \nonumber
\end{align}
Note that $W$ is a moment function on $\sZ$. Indeed, we have
\begin{align}
\lim_{n\rightarrow\infty} \inf_{z \in K_n^c} W(z) = \infty. \nonumber
\end{align}
We also have
\begin{align}
\sup_{a \in \sA} \int_{\sZ} W(z') \eta_t(dz'|z,a) \leq \alpha W(z), \text{ } \text{for all $t\geq0$}. \nonumber
\end{align}
Moreover, by \citet[Theorem 3.6]{FeKaZg16}, $\eta_t$ is weakly continuous in $(z,a)$ for all $t\geq0$. Therefore, the belief-state MDP satisfies the following conditions under Assumption~1.

\begin{itemize}
\item [(i)] The cost functions $\{C_t\}$ are bounded and continuous.
\item [(ii)] The stochastic kernels $\{\eta_t\}$ are weakly continuous.
\item [(iii)] $\sA$ is compact and $\sZ$ is $\sigma$-compact.
\item [(iv)] There exist a constant $\alpha \ge 0$  and a lower semi-continuous moment function $W: \sZ \rightarrow [0,\infty)$  such that
\begin{align}
\sup_{a \in \sA} \int_{\sZ} W(y) \eta_t(dy|z,a) \leq \alpha W(z), \text{ } \text{for all $t\geq0$.} \nonumber
\end{align}
\item [(v)] The initial probability measure $\delta_{\mu_0}$ satisfies
\begin{align}
W(\delta_{\mu_0}) \eqqcolon M < \infty. \nonumber
\end{align}
\end{itemize}

\smallskip

Our approach to prove Theorem~\ref{thm:MFE} can be summarized as follows: (i) first, we lift the partially observed stochastic control problem, that a generic agent is faced with, to a fully observed stochastic control problem (i.e., belief-state MDP) using the above mentioned results; (ii) we prove that the state transition probabilities of  the belief-state MDP are continuous with respect to state measure flow of the original partially observed problem, (iii) finally, we use the technique in our paper \citet{SaBaRa17}, which is developed to show the existence of mean-field equilibria for fully observed mean-field games, to finish the proof. The key step in our approach is (ii) in which we mimic the elegant proof technique that is established by \citet{FeKaZg16} to prove the weak continuity of the transition probabilities of fully observed reduction of partially observed stochastic control problems.

We are now ready to prove Theorem~\ref{thm:MFE}. Define the mapping $\sB: \P(\sZ) \rightarrow \P(\sX)$ as follows:
\begin{align}
\sB(\nu)(\,\cdot\,) = \int_{\sZ} z(\,\cdot\,) \text{ } \nu(dz). \nonumber
\end{align}
In other words, $\sB(\nu)$ is the so-called `barycenter' of $\nu$, see, e.g., \citet{Phe01}. Using this definition, for any $\bnu \in \P(\sZ \times \sA)^{\infty}$, we define the measure flow $\bmu^{\bnu} \in \P(\sX)^{\infty}$ as follows:
\begin{align}
\bmu^{\bnu} = \bigl(\sB(\nu_{t,1})\bigr)_{t\geq0}, \nonumber
\end{align}
where for any $\nu \in \P(\sZ \times \sA)$, we let $\nu_1$ denote the marginal of $\nu$ on $\sZ$; that is, $\nu_1(\,\cdot\,) \coloneqq \nu(\,\cdot\, \times \sA)$. Let $\{\eta_t^{\bnu}\}_{t\geq0}$ and $\{C_t^{\bnu}\}_{t\geq0}$ be, respectively, the transition probabilities and one-stage cost functions of belief-state MDP induced by the measure flow $\bmu^{\bnu}$. We let $J_{*,t}^{\bnu}: \sZ \rightarrow [0,\infty)$ denote the discounted cost value function at time $t$ of this belief-state MDP; that is,
\begin{align}
J_{*,t}^{\bnu}(z) \coloneqq \inf_{\varphi \in \Phi} E^{\varphi} \biggl[ \sum_{k=t}^{\infty} \beta^{k-t} C_k^{\bnu}(z(k),a(k)) \bigg| z(t) = z \biggr]. \nonumber
\end{align}
Let $J_{*}^{\bnu} \coloneqq \bigl( J^{\bnu}_{*,t}\bigr)_{t\geq0}$. To prove the existence of a mean-field equilibrium, we use the technique in our previous paper \citet{SaBaRa17} adopted from \citet{JoRo88}, which enables us to transform the fixed point equation $\pi \in \Psi(\Lambda(\pi))$ characterizing the mean-field equilibrium into a fixed point equation of a set-valued mapping from the set of state-action measure flows $\P(\sZ \times \sA)^{\infty}$ into itself. Then, using Kakutani's fixed point theorem (\citet[Corollary 17.55]{AlBo06}), we deduce the existence of a mean-field equilibrium.

\begin{remark}\label{remark1}
We note that the technique used here to prove the existence of a mean-field equilibrium is very similar to the one in our previous paper \citet{SaBaRa17}, in which we have studied fully-observed version of the same problem. However, there is a crucial difference in the details between this problem and the fully-observed one. In the fully-observed case, the transition probability is given by $p(\,\cdot\,|x,a,\mu_t)$ from which we can immediately deduce the continuity of the transition probability with respect to state-measure flow. However, in the partially-observed case, we do not have such an explicit analytical expression describing the relation between $\eta_t^{\bnu}$ and $\bmu^{\bnu}$ from which we can deduce the same continuity result. Hence, we need to prove this highly non-trivial  statement (unlike the fully-observed case) in order to use the technique in \citet{SaBaRa17}. Indeed, this is the key step here (step (ii) above) to prove the existence of a mean-field equilibrium. After we prove this result, the rest of the proof follows the same steps as in \citet{SaBaRa17}. However, for the sake of completeness, we give the full details of the proof, since we are using quite different notation as a result of fully-observed reduction of the original problem and the dependence of the transition probability of the belief-state MDP on the state-measure flow is not explicit as in the fully-observed case.\Halmos \end{remark}

As we mentioned above, we first transform the fixed point equation $\pi \in \Psi(\Lambda(\pi))$  into a fixed point equation of a set-valued mapping from $\P(\sZ \times \sA)^{\infty}$ into itself. To that end, we define the product space
$\C \coloneqq C_{\beta}(\sZ)^{\infty}$ in which $J_{*}^{\bnu}$ is an element. Here, $C_{\beta}(\sZ) \coloneqq \{u \in C_b(\sZ): \|u\| \leq \frac{\|c\|}{1-\beta}\}$. Moreover, we equip $\C$ with the following metric:
\begin{align}
\rho(\bu,\bv) \coloneqq \sum_{t=0}^{\infty} \sigma^{-t} \|u_t - v_t\|, \nonumber
\end{align}
where $\sigma > 0$ is chosen so that $\sigma \beta < 1$. For any  $t\geq0$, we define the \emph{Bellman optimality operator } $T_t^{\bnu}: C_{\beta}(\sZ)\rightarrow C_{\beta}(\sZ)$ as
\begin{align}
T_t^{\bnu} u(z) = \min_{a \in \sA} \biggl[ C^{\bnu}_t(z,a) + \beta \int_{\sZ} u(y) \eta^{\bnu}_t(dy|z,a) \biggr]. \nonumber
\end{align}
In the MDP theory, $T_t^{\bnu}$ gives the relation between value functions $J^{\bnu}_{*,t}$ and $J^{\bnu}_{*,t+1}$. Moreover, given $J^{\bnu}_{*,t+1}$, it characterizes the optimal policy at time $t$. It is standard to prove that $T_t^{\bnu}$ is a contraction on $C_{\beta}(\sZ)$ with modulus $\beta$ for all $t\geq0$, i.e., $\| T_t^{\bnu}(u)-T_t^{\bnu}(v) \| \le \beta \| u - v \|$ for all $u,v \in C_\beta(\sZ)$. Let us define the operator $T^{\bnu}: \C \rightarrow \C$ as
\begin{align}
\bigl( T^{\bnu} \bu \bigr)_t = T_t^{\bnu} u_{t+1}, \text{ for } t\geq0. \label{eq4}
\end{align}
It can be shown that $T^{\bnu}$ is a contraction on $\C$ with modulus $\sigma \beta$:
$$
\rho(T^{\bnu} \bu, T^{\bnu} \bv) \le \sigma\beta \cdot \rho(\bu,\bv), \qquad \forall \bu,\bv \in \C.
$$
Since $(\C,\rho)$ is a complete metric space, $T^{\bnu}$ has a unique fixed point by the Banach fixed point theorem.

The following theorem is a known result in the theory of nonhomogeneous Markov decision processes (see \citet[Theorems 14.4 and 17.1]{Hin70}). For any given $\bnu$, it characterizes $J^{\bnu}_{*}$ and the optimal policy of the belief-state MDP.

\begin{theorem}\label{theorem1}
For any $\bnu$, the collection of value functions $J^{\bnu}_{*}$ is the unique fixed point of the operator $T^{\bnu}$. Furthermore, $\varphi \in \sM$ is optimal if and only if
\begin{align}
\nu_t^{\varphi} \biggl( \biggr\{ (z,a) : C^{\bnu}_t(z,a) + \beta \int_{\sZ} J_{*,t+1}^{\bnu}(y) \eta^{\bnu}_t(dy|z,a)
= T_t^{\bnu} J_{*,t+1}^{\bnu}(z) \biggr\} \biggr) = 1, \label{eq5}
\end{align}
where $\nu_t^{\varphi} = {\cal L}\bigl( z(t),a(t) \bigr)$ under $\varphi$ and $\bnu$.
\end{theorem}

We are now ready to define the above-mentioned set-valued map from $\P(\sZ \times \sA)^{\infty}$ into itself. To that end, for any $\bnu \in \P(\sZ \times \sA)^{\infty}$, let us define the following sets:
\begin{align}
C(\bnu) &\coloneqq \biggl\{ \bnu' \in \P(\sZ \times \sA)^{\infty}: \nu'_{0,1} = \delta_{\mu_0} \text{ and }
\nu'_{t+1,1}(\,\cdot\,) = \int_{\sZ \times \sA} \eta_t^{\bnu}(\,\cdot\,|z,a) \nu_t(dz,da)\biggr\} \nonumber \\
\intertext{and}
B(\bnu) &\coloneqq \biggl\{ \bnu' \in \P(\sZ \times \sA)^{\infty}: \forall t\geq0, \text{ } \nonumber \\
&\phantom{xxxxxxxxxxxxxxxxx}\nu_t' \biggl( \biggr\{ (z,a) : C_t^{\bnu}(z,a) + \beta \int_{\sZ} J_{*,t+1}^{\bnu}(y) \eta_t^{\bnu}(dy|z,a) = T_t^{\bnu} J^{\bnu}_{*,t+1}(z) \biggr\} \biggr) = 1 \biggr\}. \nonumber
\end{align}

Note that the set $C(\bnu)$ characterizes the consistency of the mean-field term with the distribution of a generic agent, and the set $B(\bnu)$ characterizes optimality of the policy that is obtained by disintegrating the state-action measure-flow, for the mean-field term. The set-valued mapping $\Gamma: \P(\sZ \times \sA)^{\infty} \rightarrow 2^{\P(\sZ \times \sA)^{\infty}}$ is given as follows:
\begin{align}
\Gamma(\bnu) = C(\bnu) \cap B(\bnu). \nonumber
\end{align}
We say that $\bnu$ is a fixed point of $\Gamma$ if $\bnu \in \Gamma(\bnu)$. The following proposition makes the connection between mean-field equilibria and the fixed points of $\Gamma$, and so, transforms the fixed point equation $\pi \in \Psi(\Lambda(\pi))$ into the fixed point equation $\bnu \in \Gamma(\bnu)$.

\begin{proposition}\label{prop1}
Suppose that $\Gamma$ has a fixed point $\bnu = (\nu_t)_{t \ge 0}$. Construct a Markov policy $\varphi = (\varphi_t)_{t \ge 0}$ for belief-state MDP by disintegrating each $\nu_t$ as $\nu_t(dx,da) = \nu_{t,1}(dx) \varphi_t(da|x)$, and let $\bmu = (\sB(\nu_{t,1}))_{t \ge 0}$. Then the pair $(\pi^{\varphi},\bmu)$ is a mean-field equilibrium.
\end{proposition}

\proof{Proof.}
Note that, since $\bnu \in C(\bnu)$, we have $\nu_{t} = {\cal L}\bigl( z(t),a(t) \bigr)$ ($t\geq0$) for belief-state MDP under the policy $\varphi$ and the measure flow $\bmu$. Then, for any $f \in C_b(\sX)$, we have
\begin{align}
\mu_{t+1}(f) &= \sB(\nu_{t+1,1})(f) \nonumber \\
&= \int_{\sZ \times \sA} \int_{\sZ} z'(f) \eta_t^{\bnu}(dz'|z,a) \nu_t(dz,da) \nonumber \\
&= \int_{\sZ \times \sA} \biggl\{ \int_{\sX} \int_{\sX} f(y) p(dy|x,a,\mu_t) z(dx) \biggr\} \nu_t(dz,da) \nonumber \\
&= E^{\varphi} \bigl[ l_t(z(t),a(t)) \bigr] \text{  } \text{$\biggl($here $l_t(z,a) \coloneqq \int_{\sX} \int_{\sX} f(y) p(dy|x,a,\mu_t) z(dx)$$\biggr)$} \nonumber \\
&= E^{\pi^{\varphi}} \biggl[ \int_{\sX} f(y) p(dy|x(t),a(t),\mu_t) \biggr]. \label{eq:cons}
\end{align}
Since (\ref{eq:cons}) is true for all $f \in C_b(\sX)$, we have
\begin{align}
\mu_{t+1}(\,\cdot\,) = \int_{\sX \times \sA} p(\,\cdot\,|x(t),a(t),\mu_t) P^{\pi^{\varphi}}(da(t)|x(t)) \mu_t(dx(t)), \nonumber
\end{align}
where $P^{\pi^{\varphi}}(da(t)|x(t))$ denotes the conditional distribution of $a(t)$ given $x(t)$ under $\pi^{\varphi}$ and $(\mu_{\tau})_{0\leq\tau\leq t}$. Hence, $\Lambda(\pi^{\varphi}) = \bmu$.

Since $\bnu \in B(\bnu)$, the corresponding Markov policy $\varphi$ satisfies (\ref{eq5}) for $\bnu$. Therefore, by Theorem~\ref{theorem1} and the fact that $\nu_{t} = {\cal L}\bigl( z(t),a(t) \bigr)$ ($t\geq0$) for belief-state MDP under the policy $\varphi$ and the measure flow $\bmu$, $\varphi$ is optimal for belief-state MDP induced by the measure flow $\bmu$ (or, equivalently, $\bnu$). Therefore, $\pi^{\varphi} \in \Psi(\bmu)$.\Halmos
\endproof

By Proposition~\ref{prop1}, it suffices to prove that $\Gamma$ has a fixed point in order to establish the existence of a mean-field equilibrium. To prove this, Kakutani's fixed point theorem (see, e.g., \citet[Corollary 17.55]{AlBo06}),  which is a standard result to prove the existence of Nash equilibrium in classical game problems, is the right tool to use. In order to use Kakutani's fixed point theorem, the space on which the set-valued mapping is defined should be convex and compact. However, the set $\P(\sZ \times \sA)^{\infty}$ in the definition of $\Gamma$ is not compact. But, we will prove that the image of $\P(\sZ \times \sA)^{\infty}$ under $\Gamma$ is contained in a convex and compact subset of $\P(\sZ \times \sA)^{\infty}$. Hence, it is sufficient to consider this convex and compact set in the definition of $\Gamma$. To that end, for each $t\geq0$, define the set
\begin{align}
\P^t(\sZ) \coloneqq \biggl\{ \mu \in \P(\sZ): \int_{\sZ} W(z) \mu(dz) \leq \alpha^t M \biggr\}. \nonumber
\end{align}
Since $W$ is a lower semi-continuous moment function, the set $\P^t(\sZ)$ is compact with respect to the weak topology, see \citet[Proposition E.8, p. 187]{HeLa96}. Let us define
\begin{align}
\P^t(\sZ \times \sA) \coloneqq \bigl\{ \nu \in \P(\sZ \times \sA): \nu_1 \in \P^t(\sZ) \bigr\}. \nonumber
\end{align}
Since $\sA$ is compact, $\P^t(\sZ \times \sA)$ is tight. Furthermore, $\P^t(\sZ \times \sA)$ is closed with respect to the weak topology. Hence, $\P^t(\sZ \times \sA)$ is compact. Let $\Xi \coloneqq \prod_{t=0}^{\infty} \P^t(\sZ \times \sA)$, which is convex and compact with respect to the product topology.

\begin{proposition}\label{prop2}
We have $\Gamma\bigl(\P(\sZ \times \sA)^{\infty}\bigr) \coloneqq \bigl\{\bnu' : \bnu' \in \Gamma(\bnu), \text{ } \bnu \in \P(\sZ \times \sA)^{\infty} \bigr\} \subset \Xi$.
\end{proposition}

\proof{Proof.}
Fix any $\bnu \in \P(\sZ \times \sA)^{\infty}$. It is sufficient to prove that $C(\bnu) \subset \Xi$. Let $\bnu' \in C(\bnu)$. We prove by induction that $\nu'_{t,1} \in \P^t_v(\sZ)$ for all $t\geq0$. The claim trivially holds for $t=0$ as $\nu'_{0,1} = \delta_{\mu_0}$. Assume the claim holds for $t$ and consider $t+1$. We have
\begin{align}
\int_{\sZ} W(y) \nu'_{t+1,1}(dy) &= \int_{\sZ \times \sA} \int_{\sZ} W(y) \eta_{t}^{\bnu}(dy|z,a) \nu_{t}(dz,da) \nonumber \\
&\leq \int_{\sZ} \alpha W(z) \nu_{t,1}(dz) \nonumber \text{ }(\text{by (iv)}) \\
&\leq \alpha^{t+1} M \text{ }(\text{as $\nu_{t,1} \in \P^t_v(\sZ)$}). \nonumber
\end{align}
Hence, $\nu'_{t+1,1} \in \P^{t+1}_v(\sZ)$. \Halmos
\endproof

By Proposition~\ref{prop2}, it is sufficient to prove that $\Gamma$ has a fixed point $\bnu \in \Xi$. As in \citet[Theorem 1]{JoRo88}, one can prove that $C(\bnu) \cap B(\bnu) \neq \emptyset$ for any $\bnu \in \Xi$. Moreover, note that both $C(\bnu)$ and $B(\bnu)$ are convex, and thus their intersection is also convex. $\Xi$ is a convex compact subset of a locally convex topological space $\M(\sZ \times \sA)^{\infty}$, where $\M(\sZ \times \sA)$ denotes the set of all finite signed measures on $\sZ \times \sA$. Hence, in order to deduce the existence of a fixed point of $\Gamma$, we need to prove that it has a closed graph. Before stating this result, we prove the following proposition which
is a key element of the proof as mentioned earlier (i.e., step (ii)). Its proof is given in Appendix~\ref{app1}.

\begin{proposition}\label{prop:belief_conv}
Let $\bnu^{(n)} \rightarrow \bnu$ in product topology. Then, for all $t\geq0$, $\eta_t^{\bnu^{(n)}}(\,\cdot\,|z,a)$ weakly converges to $\eta_t^{\bnu}(\,\cdot\,|z,a)$ for all $(z,a) \in \sZ \times \sA$.
\end{proposition}

Using Proposition~\ref{prop:belief_conv}, we can now prove the following proposition.

\begin{proposition}\label{prop3} The graph of $\Gamma$, i.e., the set
	$$
	\Gr(\Gamma) := \left\{ (\bnu,\bxi) \in \Xi \times \Xi : \bxi \in \Gamma(\bnu)\right\},
	$$
is closed.
\end{proposition}

\proof{Proof.}
We note that modulo Proposition~\ref{prop:belief_conv}, the proof of the proposition is almost the same as the proof in \citet[Proposition 3.9]{SaBaRa17} for the full state measurement case. However, for the sake of completeness, we give the proof here.

Let $\bigl\{(\bnu^{(n)},\bxi^{(n)})\bigr\}_{n\geq1} \subset \Xi \times \Xi$ be such that $\bxi^{(n)} \in \Gamma(\bnu^{(n)})$ for all $n$ and $(\bnu^{(n)},\bxi^{(n)}) \rightarrow (\bnu,\bxi)$ as $n\rightarrow\infty$ for some $(\bnu,\bxi) \in \Xi \times \Xi$. To prove $\Gr(\Gamma)$ is closed, it is sufficient to prove that $\bxi \in \Gamma(\bnu)$.

Using Proposition~\ref{prop:belief_conv}, we first prove that $\bxi \in C(\bnu)$; that is, for all $t\geq0$, we have
\begin{align}
\xi_{t+1,1}(\,\cdot\,) = \int_{\sZ \times \sA} \eta_t^{\bnu}(\,\cdot\,|z,a) \nu_t(dz,da). \nonumber
\end{align}
For all $n$ and $t$, we have
\begin{align}
\xi^{(n)}_{t+1,1}(\,\cdot\,) = \int_{\sZ \times \sA} \eta_t^{\bnu^{(n)}}(\,\cdot\,|z,a) \nu^{(n)}_t(dz,da). \label{eq6}
\end{align}
Since $\bxi^{(n)} \rightarrow \bxi$ in $\Xi$, $\xi^{(n+1)}_{t+1} \rightarrow \xi_{t+1}$ weakly. Let $g \in C_b(\sZ)$. Then, by \citet[Theorem 3.5]{Lan81}, we have
\begin{align}
\lim_{n\rightarrow\infty} \int_{\sZ \times \sA} \int_{\sZ} g(z') \eta_t^{\bnu^{(n)}}(dz'|z,a) \nu^{(n)}_t(dz,da) =\int_{\sZ \times \sA} \int_{\sZ} g(z') \eta_t^{\bnu}(dz'|z,a)  \nu_t(dx,da) \nonumber
\end{align}
since $\bnu^{(n)}_t \rightarrow \bnu_t$ weakly and $\int_{\sZ} g(y) \eta_t^{\bnu^{(n)}}(\,\cdot\,|z,a)$ converges to $\int_{\sZ} g(y) \eta_t^{\bnu}(\,\cdot\,|z,a)$ continuously\footnote{Suppose $g$, $g_n$ ($n\geq1$) are measurable functions on metric space $\sE$. The sequence $g_n$ is said to converge to $g$ continuously if $\lim_{n\rightarrow\infty}g_n(e_n)=g(e)$ for any $e_n\rightarrow e$ where $e \in \sE$.} (see \citet[Theorem 3.5]{Lan81}). This implies that the measure in the right hand side of (\ref{eq6}) converges weakly to $\int_{\sZ \times \sA} \eta_t^{\bnu}(\,\cdot\,|z,a) \nu_t(dz,da)$. Therefore, we have
\begin{align}
\xi_{t+1,1}(\,\cdot\,) = \int_{\sZ \times \sA} \eta_t^{\bnu}(\,\cdot\,|z,a) \nu_t(dz,da), \nonumber
\end{align}
from which we conclude that $\bxi \in C(\bnu)$.

To complete the proof, it suffices to prove that $\bxi \in B(\bnu)$. To that end, for each $n$ and $t$, let us define the following functions
\begin{align}
F^{(n)}_t(z,a) &\coloneqq C_t^{\bnu^{(n)}}(z,a) + \beta \int_{\sZ} J^{\bnu^{(n)}}_{*,t+1}(y) \eta_t^{\bnu^{(n)}}(dy|z,a) \nonumber \\
\intertext{and}
F_t(z,a) &\coloneqq C_t^{\bnu}(z,a) + \beta \int_{\sZ} J^{\bnu}_{*,t+1}(y) \eta_t^{\bnu}(dy|z,a). \nonumber
\end{align}
By definition,
\begin{align}
J^{\bnu^{(n)}}_{*,t}(z) = \min_{a \in \sA} F^{(n)}_t(z,a) \text{ } \text{ and } \text{ } J^{\bnu}_{*,t}(z) = \min_{a \in \sA} F_t(z,a). \nonumber
\end{align}
Define also the following sets
\begin{align}
A_t^{(n)} &\coloneqq \bigl\{ (z,a): F^{(n)}_t(z,a) = J^{\bnu^{(n)}}_{*,t}(z) \bigr\} \nonumber \\
\intertext{and}
A_t &\coloneqq \bigl\{ (z,a): F_t(z,a) = J^{\bnu}_{*,t}(z) \bigr\}. \nonumber
\end{align}
Since $\bxi^{(n)} \in B(\bnu^{(n)})$, we have
\begin{align}
1 = \xi^{(n)}_t\bigl( A_t^{(n)} \bigr), \text{ } \text{for all $n$ and $t$}. \nonumber
\end{align}
To prove to $\bxi \in B(\bnu)$, we need to show that
\begin{align}
1 = \xi_t\bigl( A_t \bigr), \text{ } \text{for all $t$}. \nonumber
\end{align}

First note that since both $F^{(n)}_t$ and $J^{\bnu^{(n)}}_{*,t}$ are continuous, $A_t^{(n)}$ is closed. Moreover, $A_t$ is also closed as both $F_t$ and $J^{\bnu}_{*,t}$ are continuous. One can also prove as in \citet[Proposition 3.10]{SaBaRa17} that $F_t^{(n)}$ converges to $F_t$ continuously and $J^{\bnu^{(n)}}_{*,t}$ converges to $J^{\bnu}_{*,t}$ continuously, as $n\rightarrow\infty$.

For each $M\geq1$, define the closed set $B_t^M \coloneqq \bigl\{ (z,a): F_t(z,a) \geq J^{\bnu}_{*,t}(z) + \epsilon(M) \bigr\}$, where $\epsilon(M) \rightarrow 0$ as $M\rightarrow \infty$. Since both $F_t$ and $J^{\bnu}_{*,t}$ is continuous, we can choose $\{\epsilon(M)\}_{M\geq1}$ so that $\xi_t(\partial B_t^M) = 0$ for each $M$. Note that by the monotone convergence theorem, we have
\begin{align}
\xi^{(n)}_t\big(A_t^c \cap A_t^{(n)}\big) = \liminf_{M\to\infty} \xi^{(n)}_t\big(B^M_t \cap A_t^{(n)}). \nonumber
\end{align}
This implies that
\begin{align}
1 &= \limsup_{n\rightarrow\infty} \liminf_{M\rightarrow\infty} \biggl\{ \xi^{(n)}_t\big(A_t \cap A^{(n)}_t\big) + \xi^{(n)}_t\big(B^M_t \cap A_t^{(n)}\big)\biggr\} \nonumber\\
&\leq \liminf_{M\rightarrow\infty} \limsup_{n\rightarrow\infty}  \biggl\{\xi^{(n)}_t\big(A_t \cap A^{(n)}_t\big) + \xi^{(n)}_t\big(B^M_t \cap A_t^{(n)}\big)\biggr\}. \nonumber
\end{align}
\noindent For any fixed $M$, we prove that the limit of the second term in the last expression converges to zero.
To that end, we first note that $\xi^{(n)}_t$ converges weakly to $\xi_t$ as $n\rightarrow\infty$ when both measures are restricted to $B_t^M$, as $B_t^M$ is closed and $\xi_t(\partial B_t^M)=0$, see, e.g., \citet[Theorem 8.2.3]{Bog07}. Furthermore, since $F_t^{(n)}$ converges to $F_t$ continuously and $J^{\bnu^{(n)}}_{*,t}$ converges to $J^{\bnu}_{*,t}$ continuously, $1_{A^{(n)}_t \cap B^M_t}$ converges continuously to $0$, which implies by  \citet[Theorem 3.5]{Lan81} that
\begin{align*}
\limsup_{n\rightarrow\infty} \xi^{(n)}_t\big(B^M_t \cap A^{(n)}_t\big) = 0.
\end{align*}
Therefore, we obtain
\begin{align*}
1 &\leq  \limsup_{n\rightarrow\infty} \xi^{(n)}_t\big(A_t \cap A_t^{(n)}\big)\\
&\le \limsup_{n\rightarrow\infty} \xi_t^{(n)}(A_t) \nonumber \\
&\leq \xi_t(A_t), \nonumber
\end{align*}
where the last inequality follows from the Portmanteau theorem (see, e.g., \citet[Theorem 2.1]{Bil99}) and the fact that $A_t$ is closed. Hence, $\xi_t(A_t)=1$. Since $t$ is arbitrary, this is true for all $t$. This means that $\bxi \in B(\bnu)$. Therefore, $\bxi \in \Gamma(\bnu)$. \Halmos
\endproof

Recall that $\Xi$ is a compact convex subset of the locally convex topological space $\M(\sZ \times \sA)^{\infty}$. Furthermore, the graph of $\Gamma$ is closed by Proposition~\ref{prop3}, and it takes nonempty convex values. Therefore, by Kakutani's fixed point theorem (\citet[Corollary 17.55]{AlBo06}), $\Gamma$ has a fixed point. Therefore, the pair $(\pi^{\varphi},\bmu)$ is a mean field equilibrium, where $\pi^{\varphi}$ and $\bmu$ are constructed as in the statement of Proposition~\ref{prop1}. This completes the proof of Theorem~\ref{thm:MFE}.

\section{Approximation of Nash Equilibria}\label{sec4}

In this section, our aim is to show that the policy generated by the mean-field equilibrium, when adopted by each agent,  is nearly Nash equilibrium for games with sufficiently large number of agents. Let $(\pi',\bmu)$ denote the pair in the mean-field equilibrium. In addition to Assumption~\ref{as1}, we impose an additional assumption, which is stated below. To this end, let $d_{BL}$ denote the bounded Lipschitz metric (see, e.g., \citet[Proposition 11.3.2]{Dud89}) on $\P(\sX)$ that metrizes the weak topology, and define the following moduli of continuity:
\begin{align}
\omega_{p}(r) &\coloneqq \sup_{(x,a) \in \sX\times\sA} \sup_{\substack{\mu,\nu: \\ d_{BL}(\mu,\nu)\leq r}} \|p(\,\cdot\,|x,a,\mu) - p(\,\cdot\,|x,a,\nu)\|_{TV} \nonumber \\
\omega_{c}(r) &\coloneqq \sup_{(x,a) \in \sX\times\sA} \sup_{\substack{\mu,\nu: \\ d_{BL}(\mu,\nu)\leq r}} |c(x,a,\mu) - c(x,a,\nu)|. \nonumber
\end{align}

\begin{assumption}\label{as2}
\begin{itemize}
\item [ ]
\item [(a)] $\omega_p(r) \rightarrow 0$ and $\omega_c(r) \rightarrow 0$ as $r\rightarrow0$.
\item [(b)] For each $t\geq0$, $\pi_t': \sG_t \rightarrow \P(\sA)$ is deterministic; that is, $\pi_t'(\,\cdot\,|g(t)) = \delta_{f_t(g(t))}(\,\cdot\,)$ for some measurable function $f_t:\sG_t\rightarrow \sA$, and weakly continuous.
\item [(c)] The observation kernel $r(\,\cdot\,|x)$ does not depend on the mean-field term.
\end{itemize}
\end{assumption}

\begin{remark}
Note that, if the state transition probability $p$ is independent of the mean-field term, then Assumption~\ref{as2}-(a) for $\omega_p$ is always true. Indeed, in that case, we have
$\omega_p(r) = 0$ for all $r$.\Halmos
\end{remark}

\begin{remark}\label{continuity}
One way to establish Assumption~\ref{as2}-(b) is as follows. Suppose that, for $\bmu$, there exists a unique minimizer $a_z \in \sA$ of
\begin{align}
C_t^{\bmu}(z,\,\cdot\,) + \beta \int_{\sZ} J_{*,t+1}^{\bmu}(z') \eta_t^{\bmu}(dz'|z,\,\cdot\,) \eqqcolon R_t(z,\,\cdot\,), \label{unique}
\end{align}
for each $z \in \sZ$ and for all $t\geq0$. In addition, suppose that $F_t:\sZ \times \sA \times \sY \rightarrow \sZ$ ($t\geq0$) in (\ref{eq:non_filtering}) is continuous.
Note that uniqueness conditions analogous to (\ref{unique}) are quite common in mean field literature (see, e.g., \citet[Assuption 4]{GoMoSo10}, \citet[Assumption A5]{SeCa16}, \citet[Assumption H5]{HuMaCa06}, \citet[Assumption A9]{SeCa16-3}).

Under the condition of unique minimizer to (\ref{unique}), one can prove that the policy $\varphi$ in Proposition~\ref{prop1} is deterministic and weakly continuous (see \citet[Section 5]{SaBaRa17}). Indeed, fix any $t\geq0$ and consider the policy $\varphi_t$ at time $t$ in $\varphi$. By (\ref{unique}), we must have $\varphi_t(\,\cdot\,|z) = \delta_{f_t(z)}(\,\cdot\,)$ for some $f_t: \sZ\rightarrow \sA$ which minimizes $R_{t}(z,\,\cdot\,)$ of the above form; that is, $\min_{a \in \sA} R_{t}(z,a) = R_{t}(z,f_t(z))$ for all $z \in \sZ$. If $f_t$ is continuous, then $\varphi_t$ is also continuous. Hence, in order to prove the assertion, it is sufficient to prove that $f_t$ is continuous. Suppose $z_n \rightarrow z$ in $\sZ$. Note that $l_t(\,\cdot\,) = \min_{a \in \sA} R_{t}(\,\cdot\,,a)$ is continuous. Therefore, every accumulation point of the sequence $\{f_t(z_n)\}_{n\geq1}$ must be a minimizer for $R_{t}(z,\,\cdot\,)$. Since there exists a unique minimizer $f_t(z)$ of $R_{t}(z,\,\cdot\,)$, the set of all accumulation points of $\{f_t(z_n)\}_{n\geq1}$ must be $\{f_t(z)\}$. This implies that $f_t(z_n)$ converges to $f_t(z)$ since $\sA$ is compact. Hence, $f_t$ is continuous.

Recall that the mean-field equilibrium policy is given by
\begin{align}
\pi_t(\,\cdot\,|g(t)) = \varphi_t(\,\cdot\,|i(g(t))). \nonumber
\end{align}
Hence, $\pi$ is also a deterministic policy as $i$ is a deterministic function. The function $i$ can be given recursively by $F_t:\sZ \times \sA \times \sY \rightarrow \sZ$ ($t\geq0$) in (\ref{eq:non_filtering}) and the policy $\varphi$. Since $F_t$ is continuous for all $t$ and $\varphi$ is also weakly continuous, we can conclude that the mean-field policy $\pi$ is deterministic and weakly continuous. Hence, Assumption~\ref{as2}-(b) holds.

For instance, we can prove existence of a unique minimizer to (\ref{unique}) and the continuity of $F_t$ for all $t\geq0$ under the following conditions on the system components. Suppose that $\sX = \R^d$, $\sY = \R^p$, and $\sA \subset \R^m$ is convex. In addition, suppose that $p(dx'|x,a,\mu) = \varrho(x'|x,a,\mu) m(dx')$ and $r(dy|x) = \zeta(y|x) m(dy)$, where $m$ denotes the Lebesgue measure. Assume that both $\varrho$ and $\zeta$ are continuous and bounded, and $\varrho$ and $c$ are strictly convex in $a$.
Then we have
\begin{align}
H_t(dy|z,a) &= h_t(y|z,a) m(dy), \nonumber
\end{align}
where $h_t(y|z,a)$ is given by
\begin{align}
h_t(y|z,a) &= \int_{\sX} \int_{\sX} \zeta(y|x) \varrho(x|x',a,\mu_t) m(dx) z(dx'). \nonumber
\end{align}
Similarly, we have
\begin{align}
F_t(dx|z,a,y) &= f_t(dx|z,a,y) m(dx), \nonumber
\end{align}
where $f_t(x|z,a,y)$ is given by
\begin{align}
f_t(x|z,a,y) &= \frac{\int_{\sX} \zeta(y|x) \varrho(x|x',a,\mu_t) z(dx')}{\int_{\sX} \int_{\sX} \zeta(y|x) \varrho(x|x',a,\mu_t) m(dx) z(dx')} \nonumber \\
\nonumber \\
&= \frac{\int_{\sX} \zeta(y|x) \varrho(x|x',a,\mu_t) z(dx')}{h_t(y|z,a)}. \nonumber
\end{align}
One can prove that $f_t$ is continuous. Hence, $F_t$ is also continuous by \cite[Theorem 16.2]{Bil95}. To show uniqueness of a minimizer to (\ref{unique}), note that
\begin{align}
J_{*,t+1}^{\bmu}(z) &\coloneqq  \inf_{\varphi \in \Phi} E^{\varphi} \biggl[ \sum_{k=t+1}^{\infty} \beta^{k-t-1} C_k^{\bmu}(z(k),a(k)) \bigg| z(t+1) = z  \biggr] \nonumber \\
&=\inf_{\pi \in \Pi} E^{\pi} \biggl[ \sum_{k=t+1}^{\infty} \beta^{k-t-1} c(x(k),a(k),\mu_k) \bigg| x(t+1) \sim z  \biggr] = \int_{\sX} V_{*,t+1}(x) z(dx), \nonumber
\end{align}
where
\begin{align}
V_{*,t+1}(x) &\coloneqq  \inf_{\pi \in \Pi} E^{\pi} \biggl[ \sum_{k=t+1}^{\infty} \beta^{k-t-1} c(x(k),a(k),\mu_k) \bigg| x(t+1) = x  \biggr]. \nonumber
\end{align}
Hence, for any $a \in \sA$, (\ref{unique}) can be written as
\begin{align}
\int_{\sX} c(x,a,\mu_t) z(dx)& + \beta \int_{\sY} \int_{\sX} V_{*,t+1}(x) F_t(z,a,y)(dx) H_t(dy|z,a) \nonumber \\
&= \int_{\sX} c(x,a,\mu_t) z(dx) + \beta \int_{\sY} \int_{\sX} V_{*,t+1}(x) f_t(x|z,a,y) h_t(y|z,a) m(dx) m(dy)  \nonumber \\
&= \int_{\sX} c(x,a,\mu_t) z(dx) + \beta \int_{\sY} \int_{\sX} V_{*,t+1}(x) \biggl(\int_{\sX} \zeta(y|x) \varrho(x|x',a,\mu_t) z(dx')\biggr) m(dx) m(dy) \nonumber \\
&= \int_{\sX} c(x,a,\mu_t) z(dx) + \beta \int_{\sX} \int_{\sX} V_{*,t+1}(x)  \varrho(x|x',a,\mu_t) m(dx) z(dx'). \nonumber
\end{align}
Since $c$ and $\varrho$ are strictly convex in $a$, the last expression is also strictly convex in $a$. Hence, there exists a unique minimizer $a_z \in \sA$ for (\ref{unique}).\halmos
\end{remark}

For $t\geq0$, let $\sY^{t+1} \coloneqq \prod_{k=0}^t \sY$. Then, for each $t\geq1$, define $\tilde{f}_t:\sY^{t+1}\rightarrow\sA$ as
\begin{align}
\tilde{f}_t(y(t),\ldots,y(0)) \coloneqq f_t\bigl(y(t),\ldots,y(0),\tilde{f}_{t-1}(y(t-1),\ldots,y(0)),\ldots,\tilde{f}_0(y(0))\bigr), \nonumber
\end{align}
where $\tilde{f}_0 = f_0$. Let $\pi_t(\,\cdot\,|y(t),\ldots,y(0)) = \delta_{\tilde{f}_t(y(t),\ldots,y(0))}(\,\cdot\,)$. Note that $\pi_t$ is a weakly continuous stochastic kernel on $\sA$ given $\sY^{t+1}$. Indeed, $\pi$ and $\pi'$ are equivalent because, for all $t\geq0$, we have
\begin{align}
P^{\pi'}\bigl(a(t) \in \,\cdot\,|g(t)\bigr) &= P^{\pi'}\bigl(a(t) \in \,\cdot\,|y(t),\ldots,y(0)\bigr) \nonumber \\
&= P^{\pi}\bigl(a(t) \in \,\cdot\,|y(t),\ldots,y(0)\bigr). \nonumber
\end{align}
Hence, $(\pi,\bmu)$ is also a mean-field equilibrium. In the sequel, we use $(\pi,\bmu)$ to prove the approximation result. The reason for passing from $f_t$ to $\tilde{f}_t$ is that the latter policy becomes Markov in the equivalent game model that will be introduced in the proof of Theorem~\ref{appr-thm}. Thus, we can use the proof technique in our previous paper (\citet{SaBaRa17}) to show the existence of an approximate Nash equilibrium.

Before stating and proving the main result of this section on approximate Nash equilibrium property of the mean field equilibrium for the finite population case (Theorem~\ref{appr-thm}), we discuss (and establish under some conditions) the uniqueness of the mean-field equilibrium. This entails a monotonicity condition similar to the monotonicity condition introduced by \citet{LaLi07}.
\begin{assumption}\label{uniqueness-mfg}
\begin{itemize}
\item[ ]
\item[(U1)] Uniqueness condition in (\ref{unique}) holds for any mean-field equilibrium.
\item[(U2)] The state transition probability $p$ and the observation kernel $r$ do not depend on the mean-field term.
\item[(U3)] For any $\bnu$ and $\bmu$ in $\Xi$, we have the following monotonicity condition:
\begin{align}
\sum_{t=0}^{\infty} \beta^t \int_{\sZ \times \sA} \biggl( C_t^{\bmu}(z(t),a(t)) - C_t^{\bnu}(z(t),a(t)) \biggr) d\bigl(\mu_t-\nu_t\bigr)(z(t),a(t)) \geq 0. \label{monoton}
\end{align}
\end{itemize}
\end{assumption}
We note that Assumption~\ref{uniqueness-mfg} is exactly the discrete-time counterpart of the Assumption~(U) introduced in \citet{Lac15,CaLa2015}. Recall that Assumption~\ref{uniqueness-mfg}-(U1) is true under the strict convexity assumptions introduced in Remark~\ref{continuity}.

\begin{theorem}\label{unique-mfg}
Under Assumption~\ref{uniqueness-mfg}, there exists at most one solution of the mean-field equilibrium. Furthermore, if Assumption~\ref{as1} holds, then there exists a unique mean-field equilibrium.
\end{theorem}

\proof{Proof.}
Suppose that $(\pi^{\bmu},\bmu)$ and $(\pi^{\bnu},\bnu)$ are two distinct mean-field equilibria. Note that, under Assumption~\ref{uniqueness-mfg}-(U1), the policies in mean-field equilibria are deterministic; that is, $\pi^{\bmu} = \{f_t^{\bmu}\}$ and $\pi^{\bnu} = \{f_t^{\bnu}\}$, and they are unique optimal deterministic control policies given the measure flows $\bmu$ and $\bnu$. In addition, by Assumption~\ref{uniqueness-mfg}-(U2), the transition probability $\eta$ in fully-observed reduction of the POMDP in mean-field equilibrium does not depend on the mean-field term. Hence, by Assumption~\ref{uniqueness-mfg}-(U3), we have the following inequality
\begin{align}
J_{\bmu}(\pi^{\bmu}) + J_{\bnu}(\pi^{\bnu}) \geq J_{\bmu}(\pi^{\bnu}) + J_{\bnu}(\pi^{\bmu}). \nonumber
\end{align}
Since $J_{\bmu}(\pi^{\bmu}) = \inf_{\pi} J_{\bmu}(\pi)$ and $J_{\bnu}(\pi^{\bnu}) = \inf_{\pi} J_{\bnu}(\pi)$, we have
\begin{align}
J_{\bmu}(\pi^{\bmu}) = J_{\bmu}(\pi^{\bnu})  \nonumber \\
\intertext{and}
J_{\bnu}(\pi^{\bnu}) = J_{\bnu}(\pi^{\bmu}). \nonumber
\end{align}
Recall that $\pi^{\bmu}$ and $\pi^{\bnu}$ are unique optimal deterministic control policies given the measure flows $\bmu$ and $\bnu$. Therefore, $\pi^{\bmu}=\pi^{\bnu}$, and so, $\bnu = \bmu$ since the transition probability $\eta$ in fully-observed problem does not depend on mean-field term. This completes the proof.\Halmos
\endproof

The following theorem is the main result of this section, which states that the policy ${\boldsymbol \pi}^{(N)} = (\pi,\ldots,\pi)$, where $\pi$ is repeated $N$ times, is an $\varepsilon$-Nash equilibrium for sufficiently large $N$.

\begin{theorem}\label{appr-thm}
For any $\varepsilon>0$, there exists $N(\varepsilon)$ such that for $N\geq N(\varepsilon)$, the policy ${\boldsymbol \pi}^{(N)}$ is an $\varepsilon$-Nash equilibrium for the game with $N$ agents.
\end{theorem}

\subsection*{Proof of Theorem~\ref{appr-thm}}\label{sec4-1}

Note that the policy $\pi$ in the mean-field equilibrium is not necessarily Markovian, which makes the joint process of the state, observation, and mean-field term non-Markov. To prove Theorem~\ref{appr-thm}, we will first construct an equivalent game model whose states are the state of the original model plus the current and past observations. In this new model, the mean-field equilibrium policy automatically becomes Markov. Then, we will use the proof technique in our previous paper \citet{SaBaRa17} to show the existence of an approximate Nash equilibrium.

This new game model is specified by
\begin{align}
\biggl( \{\sS_t\}_{t\geq0}, \sA, \{P_t\}_{t\geq0}, \{C_t\}_{t\geq0}, \lambda_0 \biggr), \nonumber
\end{align}
where, for each $t\geq0$,
\begin{align}
\sS_t &= \sX \times \underbrace{\sY \times \ldots \times \sY}_{\text{$t+1$-times}} \nonumber
\end{align}
and $\sA$ are the Polish state and action spaces at time $t$, respectively. The stochastic kernel $P_t : \sS_t \times \sA \times \P(\sS_t) \to \P(\sS_{t+1})$ is defined as:
\begin{align}
&P_t\bigl(B_{t+1} \times D_{t+1} \times \ldots \times D_0 \big| s(t),a(t),\Delta_t\bigr) \nonumber \\
&\phantom{xxxxxxx}\coloneqq \int_{B_{t+1}} r(D_{t+1}|x(t+1)) \prod_{k=0}^t 1_{D_k}(y(k)) p(dx(t+1)|x(t),a(t),\Delta_{t,1}) , \nonumber
\end{align}
where $B_{t+1} \in \B(\sX)$, $D_k \in \B(\sY)$ ($k=0,\ldots,t+1$), $s(t) = (x(t),y(t),y(t-1),\ldots,y(0))$, and $\Delta_{t,1}$ is the marginal of $\Delta_t$ on $\sX$. Indeed, $P_t$ is the controlled transition probability of next state-observation pair, current observation, and past observations $\bigl(x(t+1),y(t+1),y(t),\ldots,y(0)\bigr)$ given the current state-observation pair and past observations $\bigl(x(t),y(t),y(t-1),\ldots,y(0)\bigr)$ in the original mean-field game. For each $t\geq0$, the one-stage cost function $C_t: \sS_t \times \sA \times \P(\sS_t) \rightarrow [0,\infty)$ is defined as:
\begin{align}
C_t(s(t),a(t),\Delta_t) \coloneqq c(x(t),a(t),\Delta_{t,1}). \nonumber
\end{align}
Finally, the initial measure $\lambda_0$ is given by $\lambda_0(ds(0)) \coloneqq r(dy|x) \mu_0(dx)$, where $s(0) = (x,y)$.

Suppose that Assumptions~\ref{as1} and \ref{as2} hold. For each $t\geq0$, let $d_t$ denote the bounded Lipschitz metric on $\P(\sS_t)${\footnote{The product metric on $\sS_t$ is assumed to be the sum of the metrics of the components in the product space.}}, and define the following moduli of continuity:
\begin{align}
\omega_{P_t}(r) &\coloneqq  \sup_{(s,a) \in \sS_t\times\sA} \sup_{\substack{\Delta,\Delta': \\ d_t(\Delta,\Delta')\leq r}}  \|P_t(\,\cdot\,|s,a,\Delta) - P_t(\,\cdot\,|s,a,\Delta')\|_{TV} \nonumber \\
\omega_{C_t}(r) &\coloneqq \sup_{(s,a) \in \sS_t\times\sA} \sup_{\substack{\Delta,\Delta': \\ d_t(\Delta,\Delta')\leq r}} |C_t(s,a,\Delta) - C_t(s,a,\Delta')|. \nonumber
\end{align}
Then, for each $t\geq0$, the following are satisfied:
\begin{itemize}
\item [(I)] The one-stage cost function $C_t$ is bounded and continuous.
\item [(II)] The stochastic kernel $P_t$ is weakly continuous.
\item [(III)] $\omega_{P_t}(r)\rightarrow0$ and $\omega_{C_t}(r)\rightarrow0$ as $r\rightarrow0$.
\end{itemize}
It is straightforward to prove that (I) and (II) hold since $c$ is continuous, $p$ is weakly continuous, and $r$ is continuous in total variation norm. For (III), for each $t\geq0$, fix any $(s,a) \in S_t \times \sA$ and $(\Delta,\Delta') \in \P(S_t) \times \P(S_t)$, and for any $f:\sS_{t+1} \rightarrow \R$, define $\tilde{f}(\,\cdot\,,\,\cdot\,) = f(\,\cdot\,,\,\cdot\,,y(t),\ldots,y(0))$ where $s= (x(t),y(t),\ldots,y(0))$. Then, we have
\begin{align}
&\sup_{\|f\|\leq1} \biggl| \int_{\sS_{t+1}} f(s') P_t(ds'|s,a,\Delta) - \int_{\sS_{t+1}} f(s') P_t(ds'|s,a,\Delta') \biggr| \nonumber \\
&\phantom{xxxxxx}= \sup_{\|f\|\leq1} \biggl| \int_{\sX \times \sY} \tilde{f}(x'(t+1),y'(t+1)) r(dy'(t+1)|x'(t+1)) p(dx'(t+1)|x(t),a,\Delta_{1}) \nonumber \\
&\phantom{xxxxxxxxxxxxxxxx} - \int_{\sX \times \sY} \tilde{f}(x'(t+1),y'(t+1)) r(dy'(t+1)|x'(t+1)) p(dx'(t+1)|x(t),a,\Delta'_{1}) \biggr| \nonumber \\
&\phantom{xxxxxx}\leq \sup_{\|g\|\leq1} \biggl| \int_{\sX} g(x') p(dx'|x(t),a,\Delta_{1}) -  \int_{\sX} g(x') p(dx'|x(t),a,\Delta'_{1}) \biggr| \nonumber \\
&\phantom{xxxxxx}\leq \omega_p(d_{BL}(\Delta_{1},\Delta'_{1})). \nonumber
\end{align}
Since $(s,a) \in S_t \times \sA$ is arbitrary and $d_{BL}(\Delta_{1},\Delta'_{1}) \leq d_{t}(\Delta,\Delta')$, we have $\omega_{P_t}(r)\rightarrow 0$ as $r\rightarrow0$ by Assumption~\ref{as2}-(a). Similarly, we can also prove that $\omega_{C_t}(r)\rightarrow0$ as $r\rightarrow0$.

Recall the set of policies $\Pi$ in the original mean-field game. Let $\tilde{\Pi}$ be the set of policies in $\Pi$ which only use the observations; that is, $\pi \in \tilde{\Pi}$ if $\pi_t:\sY^{t+1} \rightarrow \P(\sA)$ for each $t\geq0$. Note that $\tilde{\Pi}$ is a subset of the set of Markov policies in the new model. For any measure flow ${\boldsymbol \Delta} = (\Delta_t)_{t\geq0}$, where $\Delta_t \in \P(\sS_t)$, we denote by $\hat{J}_{{\boldsymbol \Delta}}(\pi)$ the infinite-horizon discounted-cost of the policy $\pi \in \tilde{\Pi}$ in this new model.

Similar to Section~\ref{sec2}, we also define the corresponding $N$ agent game as follows.
We have the Polish state spaces $\{\sS_t\}_{t\geq0}$ and action space $\sA$. For every $t \in \{0,1,2,\ldots\}$ and every $i \in \{1,2,\ldots,N\}$, let $s^N_i(t) \in \sS_t$ and $a^N_i(t) \in \sA$ denote the state and the action of Agent~$i$ at time $t$, and let
\begin{align}
\Delta_t^{(N)}(\,\cdot\,) \coloneqq \frac{1}{N} \sum_{i=1}^N \delta_{s_i^N(t)}(\,\cdot\,) \in \P(\sS_t) \nonumber
\end{align}
denote the empirical distribution of the state configuration at time $t$. The initial states $s^N_i(0)$ are independent and identically distributed according to $\lambda_0$, and, for each $t \ge 0$, the next-state configuration $(s^N_1(t+1),\ldots,s^N_N(t+1))$ is generated at random according to the probability laws
\begin{align}
&\prod^N_{i=1} P_{t}\big(ds^N_i(t+1)\big|s^N_i(t),a^N_i(t),\Delta^{(N)}_t\big). \nonumber 
\end{align}

Recall that $\tilde{\Pi}_i$ denotes the set of policies that only use local observations for Agent $i$ in the original game. Note that $\tilde{\Pi}_i$ is an admissible class of policies for the new model. Indeed, policies in $\tilde{\Pi}_i$ are Markov for this new model since they partly use the state information. We let $\tilde{\Pi}_i^c$ denote the set of all policies in $\tilde{\Pi}_i$ for Agent~$i$ that are weakly continuous; that is, $\pi=\{\pi_t\}\in \tilde{\Pi}_i^c$ if for all $t\geq0$, $\pi_t: \sY^{t+1} \rightarrow \P(\sA)$ is continuous when $\P(\sA)$ is endowed with the weak topology.

For Agent~$i$, the infinite-horizon discounted cost under the initial distribution $\lambda_0$ and $N$-tuple of policies ${\boldsymbol \pi}^{(N)} \in \tilde{{\bf \Pi}}^{(N)}$ is denoted as $\hat{J}_i^{(N)}({\boldsymbol \pi}^{(N)})$.

The following proposition makes the connection between this new model and the original model.

\begin{proposition}\label{app-prop1}
For any $N\geq1$, ${\boldsymbol \pi}^{(N)} \in \tilde{{\bf \Pi}}^{(N)}$, and $i=1,\ldots,N$, we have $\hat{J}_i({\boldsymbol \pi}^{(N)}) = J_i({\boldsymbol \pi}^{(N)})$. Similarly, for any $\pi \in \tilde{\Pi}$ and measure flow ${\boldsymbol \Delta}$, we have $\hat{J}_{\boldsymbol \Delta}(\pi) = J_{\bmu}(\pi)$ where $\bmu = (\Delta_{t,1})_{t\geq0}$.
\end{proposition}

\proof{Proof.}
The proof of the proposition is given in Appendix~\ref{app3}.\Halmos
\endproof

By Proposition~\ref{app-prop1}, in the remainder of this section we consider the new game model in place of the original one. Then, we use the same technique as in our previous paper \citet{SaBaRa17} to prove the approximation result since the policy in the mean-field equilibrium is Markov for this new model. However, as the state space in this new model is expanding at each time step, there will be some differences between the current proof and the proof in \citet[Section 4]{SaBaRa17}. Therefore, for the sake of completeness, we give the full details of the proof.

Define the measure flow ${\boldsymbol \Delta} = (\Delta_t)_{t\geq0}$ as follows: $\Delta_t = {\cal L}(x(t),y(t),\ldots,y(0))$, where ${\cal L}(x(t),y(t),\ldots,y(0))$ denotes the probability law of
$(x(t),y(t),\ldots,y(0))$ in the original mean-field game under the policy $\pi$ in the mean-field equilibrium.
For each $t\geq0$, define the stochastic kernel $P_t^{\pi}(\,\cdot\,|s,\Delta)$ on $\sS_{t+1}$ given $\sS_{t} \times \P(\sS_{t})$ as
\begin{align}
P_t^{\pi}(\,\cdot\,|s,\Delta) \coloneqq \int_{\sA} P_t(\,\cdot\,|s,a,\Delta) \pi_t(da|s). \nonumber
\end{align}
Since $\pi_t$ is assumed to be weakly continuous, $P_t^{\pi}(\,\cdot\,|s,\Delta)$ is also weakly continuous in $(s,\Delta)$. In the sequel, to ease the notation, we will also write $P_t^{\pi}(\,\cdot\,|s,\Delta)$ as $P_{t,\Delta}^{\pi}(\,\cdot\,|s)$.

\begin{lemma}\label{app-lemma1}
Measure flow ${\boldsymbol \Delta}$ satisfies
\begin{align}
\Delta_{t+1}(\,\cdot\,) &= \int_{\sS_t} P_{t}^{\pi}(\,\cdot\,|s,\Delta_t) \Delta_t(ds) \nonumber \\
&= \Delta_t P_{t,\Delta_t}^{\pi}(\,\cdot\,). \nonumber
\end{align}
\end{lemma}

\proof{Proof.}
The proof of the lemma is given in Appendix~\ref{app4}.\Halmos
\endproof

For each $N\geq1$, let $\bigl\{s_i^{N}(t)\bigr\}_{1\leq i\leq N}$ denote the state configuration at time $t$ in the $N$-person game under the policy ${\boldsymbol \pi}^{(N)} = \{\pi,\pi,\ldots,\pi\}$. Define the empirical distribution
\begin{align}
\Delta_t^{(N)}(\,\cdot\,) \coloneqq \frac{1}{N} \sum_{i=1}^N \delta_{s_i^{N}(t)}(\,\cdot\,). \nonumber
\end{align}

\begin{proposition}\label{prop5}
For all $t\geq0$, we have
\begin{align}
{\cal L}(\Delta_t^{(N)}) \rightarrow \delta_{\Delta_t} \nonumber
\end{align}
weakly in $\P(\P(\sS_t))$, as $N\rightarrow\infty$.
\end{proposition}

\proof{Proof.}
It is known that weak topology on $\P(\sS_t)$ can be metrized using the following metric:
\begin{align}
\rho(\mu,\nu) \coloneqq \sum_{m=1}^{\infty} 2^{-(m+1)} | \mu(f_m) - \nu(f_m) |, \nonumber
\end{align}
where $\{f_m\}_{m\geq1}$ is an appropriate sequence of real continuous and bounded functions on $\sS_t$ such that $\|f_m\| \leq 1$ for all $m\geq1$ (see \citet[Theorem 6.6, p. 47]{Par67}). Define the Wasserstein distance of order 1 on the set of probability measures $\P(\P(\sS_t))$ as follows (see \citet[Definition 6.1]{Vil09}):
\begin{align}
W_1(\Phi,\Psi) \coloneqq \inf \bigl\{ E[\rho(X,Y)]: {\cal L}(X) = \Phi \text{ and } {\cal L}(Y) = \Psi \bigr\}. \nonumber
\end{align}
Note that since $\delta_{\Delta_t}$ is a Dirac measure, we have
\begin{align}
W_1({\cal L}(\Delta_t^{(N)}),\delta_{\Delta_t}) &= \bigl\{ E[\rho(X,Y)]: {\cal L}(X) = {\cal L}(\Delta_t^{(N)}) \text{ and } {\cal L}(Y) = \delta_{\Delta_t} \bigr\} \nonumber \\
&= E\biggl[ \sum_{m=1}^{\infty} 2^{-(m+1)} | \Delta_t^{(N)}(f_m) - \Delta_t(f_m) | \biggr]. \nonumber
\end{align}
Since convergence in $W_1$ distance implies weak convergence (see \citet[Theorem 6.9]{Vil09}), it suffices to prove that
\begin{align}
\lim_{N\rightarrow\infty} E\bigl[|\Delta_t^{(N)}(f) - \Delta_t(f)|\bigr] = 0 \nonumber
\end{align}
for any $f \in C_b(\sS_t)$ and for all $t$. We prove this by induction on $t$.

As $\{s_i^N(0)\}_{1\leq i\leq N} \sim \lambda_0^{\otimes N}$, the claim is true for $t=0$. We suppose that the claim holds for $t$ and consider $t+1$. Fix any $g \in C_b(\sS_{t+1})$. Then, we have
\begin{align}
|\Delta_{t+1}^{(N)}(g) - \Delta_{t+1}(g)| &\leq |\Delta_{t+1}^{(N)}(g) - \Delta_{t}^{(N)} P^{\pi}_{t,\Delta_t^{(N)}}(g)| + |\Delta_t^{(N)} P^{\pi}_{t,\Delta_t^{(N)}}(g) - \Delta_t P^{\pi}_{t,\Delta_t}(g) |. \label{eq8}
\end{align}
We first prove that the expectation of the second term on the right-hand side (RHS) of (\ref{eq8}) converges to $0$ as $N\rightarrow\infty$. To that end, define $F: \P(\sS_{t}) \rightarrow \R$ as
\begin{align}
F(\Delta) = \Delta P^{\pi}_{t,\Delta}(g) \coloneqq \int_{\sS_t} \int_{\sS_{t+1}} g(s') P^{\pi}_t(ds'|s,\Delta) \Delta(ds). \nonumber
\end{align}

One can prove that $F \in C_b(\P(\sS_t))$. Indeed, suppose that $\Delta_n$ converges to $\Delta$. Let us define
\begin{align}
l_n(s) &= \int_{\sS_{t+1}} g(s') P^{\pi}_t(ds'|s,\Delta_n) \nonumber \\
\intertext{and}
l(s) &= \int_{\sS_{t+1}} g(s') P^{\pi}_t(ds'|s,\Delta). \nonumber
\end{align}
Since $P^{\pi}_t$ is weakly continuous, one can prove that $l_n$ converges to $l$ continuously; that is, if $s_n$ converges to $s$, then $l_n(s_n) \rightarrow l(s)$. By \citet[Theorem 3.5]{Lan81}, we have $F(\Delta_n) \rightarrow F(\Delta)$, and so, $F \in C_b(\P(\sS_t))$. This implies that the expectation of the second term on the RHS of (\ref{eq8}) converges to zero as ${\cal L}(\Delta_t^{(N)}) \rightarrow \Delta_t$ weakly, by the induction hypothesis.

Now, let us write the expectation of the first term on the RHS of (\ref{eq8}) as
\begin{align}
E\biggl[ E\biggl[ |\Delta_{t+1}^{(N)}(g) - \Delta_t^{(N)} P^{\pi}_{t,\Delta_t^{(N)}}(g)| \biggr| s_1^N(t),\ldots,s_N^N(t) \biggr] \biggr]. \nonumber
\end{align}
Then, by \citet[Lemma A.2]{BuMa14}, we have
\begin{align}
E\biggl[ |\Delta_{t+1}^{(N)}(g) - \Delta_t^{(N)} P^{\pi}_{t,\Delta_t^{(N)}}(g)| \biggr| s_1^N(t),\ldots,s_N^N(t) \biggr] \leq 2 \frac{\|g\|}{\sqrt{N}}. \nonumber
\end{align}
Therefore, the expectation of the first term on the RHS of (\ref{eq8}) also converges to zero as $N\rightarrow\infty$. Since $g$ is arbitrary, this completes the proof.\Halmos
\endproof

Proposition~\ref{prop5} essentially says that in the infinite-population limit, the empirical distribution of the states under the mean-field policy converges to the deterministic measure flow ${\boldsymbol \Delta}$. Since the transition probabilities $P_t(\,\cdot\,|s,a,\Delta)$ are continuous in $\Delta$, the evolution of the state of a generic agent in the finite-agent game with sufficiently many agents and the evolution of the state in the mean-field game under policies $\bpi^{(N)} = (\pi,\ldots,\pi)$ and $\pi$, respectively, should therefore be close. Hence, the distributions of the states in each problem should also be close, from which we obtain the following result.

\begin{proposition}\label{prop6}
We have
\begin{align}
\lim_{N\rightarrow\infty} \hat{J}_1^{(N)}({\boldsymbol \pi}^{(N)}) = \hat{J}_{{\boldsymbol \Delta}}(\pi) = \inf_{\pi' \in \Pi} \hat{J}_{{\boldsymbol \Delta}}(\pi'). \nonumber
\end{align}
\end{proposition}

\proof{Proof.}
For each $t\geq0$, let us define
\begin{align}
C_{\pi_t}(s,\Delta) \coloneqq \int_{\sA} C_t(s,a,\Delta) \pi_t(da|s). \nonumber
\end{align}
Since for any permutation $\sigma$ of $\{1,\ldots,N\}$ we have
\begin{align}
{\cal L}\bigl(s_1^N(t),\ldots,s_N^N(t),\Delta_t^{(N)}\bigr) = {\cal L}\bigl(s_{\sigma(1)}^N(t),\ldots,s_{\sigma(N)}^N(t),\Delta_t^{(N)}\bigr), \nonumber
\end{align}
the cost function at time $t$ can be written as
\begin{align}
E\bigl[ C_t(s_1^N(t),a_1^N(t),\Delta_t^{(N)}) \bigr] &= \frac{1}{N} \sum_{i=1}^N E\bigl[ C_t(s_i^N(t),a_i^N(t),\Delta_t^{(N)}) \bigr] \nonumber \\
&= E\bigl[ \Delta_t^{(N)}\bigl(C_{\pi_t}(s,\Delta_t^{(N)})\bigr) \bigr]. \nonumber
\end{align}
Let $F: \P(\sS_t) \rightarrow \R$ be defined as
\begin{align}
F(\Delta) \coloneqq \int_{\sS_t} C_{\pi_t}(s,\Delta) \Delta(ds). \nonumber
\end{align}
One can show that $F \in C_b(\P(\sS_t))$ as $\pi_t$ is weakly continuous. Hence, by Proposition~\ref{prop5} we obtain
\begin{align}
\lim_{N\rightarrow\infty} E\bigl[ C_t(s_1^N(t),a_1^N(t),\Delta_t^{(N)}) \bigr] &= \lim_{N\rightarrow\infty} E\bigl[ \Delta_t^{(N)}\bigl(C_{\pi_t}(s,\Delta_t^{(N)})\bigr) \bigr] \nonumber \\
&= \lim_{N\rightarrow\infty} E[F(\Delta_t^{(N)})] \nonumber \\
&= F(\Delta_t) \nonumber \\
&= \Delta_t(C_{\pi_t}(\,\cdot\,,\Delta_t)). \label{eq9}
\end{align}
Note that by Lemma~\ref{app-lemma1}, the discounted cost in the mean-field game can be written as
\begin{align}
\hat{J}_{{\boldsymbol \Delta}}(\pi) = \sum_{t=0}^{\infty} \beta^t \Delta_t(C_{\pi_t}(\,\cdot\,,\Delta_t)). \nonumber
\end{align}
Therefore, by (\ref{eq9}) and the dominated convergence theorem, we obtain
\begin{align}
\lim_{N\rightarrow\infty} \hat{J}_1^{(N)}({\boldsymbol \pi}^{(N)}) = \hat{J}_{{\boldsymbol \Delta}}(\pi), \nonumber
\end{align}
which completes the proof.\Halmos
\endproof

In order to prove the approximation result, we have to prove that if the policy of some agent deviates from the mean-field equilibrium policy, then the corresponding cost of this agent should be close to the cost in the mean-field limit as in Proposition~\ref{prop6}, for $N$ sufficiently large. However, note that this agent can choose different policies for each $N$ in place of the mean-field equilibrium policy. Since the transition probabilities and the one-stage cost functions are the same for all agents in the game model, it is sufficient to change the policy of Agent~$1$ for each $N$. To that end, let $\{\tpi^{(N)}\}_{N\geq1} \subset \tilde{\Pi}_1^c$ be an arbitrary sequence of policies for Agent~$1$; that is, for each $N\geq1$ and $t\geq0$, $\tpi_t^{(N)}: \sY^{t+1} \rightarrow \P(\sA)$ is weakly continuous. For each $N\geq1$, let $\bigl\{\ts_i^N(t)\bigr\}_{1\leq i \leq N}$ be the collection of states in the $N$-person game under the policy $\tilde{{\boldsymbol \pi}}^{(N)} \coloneqq \{\tpi^{(N)},\pi,\ldots,\pi\}$. Define
\begin{align}
\tilde{\Delta}_t^{(N)}(\,\cdot\,) \coloneqq \frac{1}{N} \sum_{i=1}^N \delta_{\ts_i^{(N)}(t)}(\,\cdot\,). \nonumber
\end{align}
The following result states that, in the infinite-population limit, the law of the empirical distribution of the states at each time $t$ is insensitive to local deviations from the mean-field equilibrium policy.

\begin{proposition}\label{prop8}
For all $t\geq0$, we have
\begin{align}
{\cal L}(\tilde{\Delta}_t^{(N)}) \rightarrow \delta_{\Delta_t} \nonumber
\end{align}
weakly $\P(\P(\sS_t))$, as $N \rightarrow \infty$.
\end{proposition}

\proof{Proof.}
The proof can be done by slightly modifying the proof of Proposition~\ref{prop5}, and therefore will not be included here. See the proof of \citet[Proposition 4.6]{SaBaRa17}.\Halmos
\endproof

For each $N\geq1$, let $\{\hs^N(t)\}_{t\geq0}$ denote the state trajectory of the mean-field game under policy $\tpi^{(N)}$; that is, $\hs^N(t)$ evolves as follows:
\begin{align}
\hs^N(0) \sim \lambda_0 \text{ and } \hs^N(t+1) \sim P^{\tpi^{(N)}}_{t,\Delta_t}(\,\cdot\,|\hs^N(t)). \nonumber
\end{align}
Recall that the cost function of this mean-field game is given by
\begin{align}
\hat{J}_{{\boldsymbol \Delta}}(\tpi^{(N)}) = \sum_{t=0}^{\infty} \beta^t E\bigl[ C_t(\hs^N(t),\hat{a}^N(t),\Delta_t)\bigr], \label{nneq1}
\end{align}
where the action configuration at each time $t\geq0$ is generated according to the probability law
\begin{align}
\tilde{\pi}^{(N)}_t(d\hat{a}^N(t)|\hs^N(t)) = \tilde{\pi}^{(N)}_t(d\hat{a}^N(t)|\hat{y}^N(t),\ldots,\hat{y}^N(0)).\nonumber
\end{align}

\begin{proposition}\label{prop9}
For any $t\geq0$, we have
\begin{align}
\lim_{N\rightarrow\infty} \bigl| {\cal L}(\ts_1^N(t),\tilde{\Delta}_t^{(N)})(T_N) - {\cal L}(\hs^N(t),\delta_{\Delta_t})(T_N) \bigr| = 0 \nonumber
\end{align}
for any sequence $\{T_N\} \subset C_b(\sS_t\times\P(\sS_t))$ such that the family $\bigl\{T_N(s,\,\cdot\,): s \in \sS_t, N\geq1)\bigr\}$ is equicontinuous and $\sup_{N\geq1}\|T_N\|<\infty$.
\end{proposition}

\proof{Proof.}
The proof of Proposition~\ref{prop9} is given in Appendix~\ref{app6}.\Halmos
\endproof

Using Proposition~\ref{prop9}, we now prove the following theorem which is a key element in the proof of Theorem~\ref{appr-thm}.

\begin{theorem}\label{theorem3}
Let $\{\tpi^{(N)}\}_{N\geq1} \subset \tilde{\Pi}_1^c$ be an arbitrary sequence of policies for Agent~$1$. Then, we have
\begin{align}
\lim_{N \rightarrow \infty} \bigl| \hat{J}_1^{(N)}(\tpi^{(N)},\pi,\ldots,\pi) - \hat{J}_{{\boldsymbol \Delta}}(\tpi^{(N)}) \bigr| = 0, \nonumber
\end{align}
where $\hat{J}_{{\boldsymbol \Delta}}(\tpi^{(N)})$ is given in (\ref{nneq1}).
\end{theorem}

\proof{Proof.}
Fix any $t\geq0$ and define
\begin{align}
T_{N,t}(s,\Delta) \coloneqq \int_{\sA} C_t(s,a,\Delta) \tpi_t^{(N)}(da|s). \nonumber
\end{align}
We first prove that $\bigl\{T_{N,t}(s,\,\cdot\,): s \in \sS_t, N\geq1\bigr\}$ satisfies the hypothesis in Proposition~\ref{prop9}. For equicontinuity, note that for any $s \in \sS_t$ and for any $(\Delta,\Delta') \in \P(\sS_t)^2$, we have
\begin{align}
\bigl| T_{N,t}(s,\Delta) - T_{N,t}(s,\Delta') \bigr| &= \biggl| \int_{\sA} C_t(s,a,\Delta) \tpi_t^{(N)}(da|s) - \int_{\sA} C_t(s,a,\Delta') \tpi_t^{(N)}(da|s) \biggr| \nonumber \\
&\leq \omega_{C_t}(d_t(\Delta,\Delta')). \nonumber
\end{align}
Since $\omega_{C_t}(r) \rightarrow 0$ as $r\rightarrow0$, the family $\bigl\{T_{N,t}(s,\,\cdot\,): s \in \sS_t, N\geq1\bigr\}$ is equicontinuous. Moreover, $\sup_{N\geq1} \|T_{N,t}\| < \infty$.

Therefore, by Proposition~\ref{prop9}, we have
\begin{align}
\lim_{N\rightarrow\infty} \biggl| E\bigl[ C_t(\ts_1^N,\tilde{a}_1^N,\tilde{\Delta}_t^{(N)}) \bigr] - E\bigl[ C_t(\hs_1^N,\hat{a}_1^N,\Delta_t) \bigr] \biggr|=0. \nonumber
\end{align}
Since $t$ is arbitrary, the above result holds for all $t\geq0$. Then the theorem follows from the dominated convergence theorem.\Halmos
\endproof

As a corollary of Propositions~\ref{app-prop1} and \ref{prop6}, and Theorem~\ref{theorem3}, we obtain the following result.

\begin{corollary}\label{cor1}
We have
\begin{align}
\lim_{N \rightarrow \infty} \hat{J}_1^{(N)}(\tpi^{(N)},\pi,\ldots,\pi)
&\geq \inf_{\pi' \in \tilde{\Pi}} \hat{J}_{{\boldsymbol \Delta}}(\pi') \nonumber \\
&= \hat{J}_{{\boldsymbol \Delta}}(\pi) \nonumber \\
&= \lim_{N \rightarrow \infty} \hat{J}_1^{(N)}(\pi,\pi,\ldots,\pi). \nonumber
\end{align}
\end{corollary}

Now, we are ready to prove the main result of this section.

\proof{Proof of Theorem~\ref{appr-thm}}
One can prove that for any policy ${\boldsymbol \pi}^{(N)} \in \tilde{{\bf \Pi}}^{(N)}$, we have
\begin{align}
\inf_{\pi^i \in \tilde{\Pi}_i} \hat{J}_i^{(N)}({\boldsymbol \pi}^{(N)}_{-i},\pi^i) = \inf_{\pi^i \in \tilde{\Pi}_i^c} \hat{J}_i^{(N)}({\boldsymbol \pi}^{(N)}_{-i},\pi^i) \nonumber
\end{align}
for each $i=1,\ldots,N$ (see the proof of \citet[Theorem 2.3]{SaBaRa17}). Hence, it is sufficient to consider weakly continuous policies in ${\bf \Pi}^{(N)}$ to establish the existence of $\varepsilon$-Nash equilibrium in the new model.

We first prove that for sufficiently large $N$, we have
\begin{align}
\hat{J}_i^{(N)}({\boldsymbol \pi}^{(N)}) &\leq \inf_{\pi^i \in \tilde{\Pi}_i^c} \hat{J}_i^{(N)}({\boldsymbol \pi}^{(N)}_{-i},\pi^i) + \varepsilon \label{eq13}
\end{align}
for each $i=1,\ldots,N$. As indicated earlier, since the transition probabilities and the one-stage cost functions are the same for all agents in the new game, it is sufficient to prove (\ref{eq13}) for Agent~$1$ only. Given $\epsilon > 0$, for each $N\geq1$, let $\tpi^{(N)} \in \tilde{\Pi}_1^c$ be such that
\begin{align}
\hat{J}_1^{(N)} (\tpi^{(N)},\pi,\ldots,\pi) < \inf_{\pi' \in \tilde{\Pi}_1^c} \hat{J}_1^{(N)} (\pi',\pi,\ldots,\pi) + \frac{\varepsilon}{3}. \nonumber
\end{align}
Then, by Corollary~\ref{cor1}, we have
\begin{align}
\lim_{N\rightarrow\infty} \hat{J}_1^{(N)} (\tpi^{(N)},\pi,\ldots,\pi) &= \lim_{N\rightarrow\infty} \hat{J}_{{\boldsymbol \Delta}}(\tpi^{(N)}) \nonumber \\
&\geq \inf_{\pi'} \hat{J}_{{\boldsymbol \Delta}}(\pi') \nonumber \\
&= \hat{J}_{{\boldsymbol \Delta}}(\pi) \nonumber \\
&= \lim_{N\rightarrow\infty} \hat{J}_1^{(N)} (\pi,\pi,\ldots,\pi). \nonumber
\end{align}
Therefore, there exists $N(\varepsilon)$ such that for $N\geq N(\varepsilon)$, we have
\begin{align}
\inf_{\pi' \in \tilde{\Pi}_1^c} \hat{J}_1^{(N)} (\pi',\pi,\ldots,\pi) + \varepsilon &> \hat{J}_1^{(N)} (\tpi^{(N)},\pi,\ldots,\pi) + \frac{2\varepsilon}{3} \nonumber \\
&\geq \hat{J}_{{\boldsymbol \Delta}}(\pi) + \frac{\varepsilon}{3} \nonumber \\
&\geq \hat{J}_1^{(N)} (\pi,\pi,\ldots,\pi). \nonumber
\end{align}
The result then follows from Proposition~\ref{app-prop1}.\Halmos
\endproof

\begin{remark}
We note that, using similar ideas, the finite-horizon cost criterion
\begin{align}
E \biggl[ \sum_{t=0}^{T} c(x(t),a(t),\mu_t)\biggr] \text{ } \text{ for some $T<\infty$},
\end{align}
can be handled with the same quantitative results. The only part that requires a verification different from the infinite-horizon case is the following result: $F_t^{(n)}$ and $J_{*,t}^{\bnu^{(n)}}$ converge continuously to $F_t$ and $J^{\bnu}_{*,t}$, respectively. Note that, in the finite-horizon case, for each $n$ and $t<T$, these functions are given by
\begin{align}
F^{(n)}_t(z,a) &= C_t^{\bnu^{(n)}}(z,a) + \int_{\sZ} J^{\bnu^{(n)}}_{*,t+1}(y) \eta^{\bnu^{(n)}}(dy|z,a), \nonumber \\
F_t(z,a) &= C_t^{\bnu}(z,a) + \int_{\sZ} J^{\bnu}_{*,t+1}(y) \eta^{\bnu}(dy|z,a), \nonumber
\end{align}
and
\begin{align}
J^{\bnu^{(n)}}_{*,t}(z) = \min_{a \in \sA} F^{(n)}_t(z,a) \text{ } \text{ and } \text{ } J^{\bnu}_{*,t}(z) = \min_{a \in \sA} F_t(z,a). \nonumber
\end{align}
Note that the discount factor $\beta$ is missing in the above equations. For $t=T$, we have $F_T^{(n)}(z,a) = C_T^{\bnu^{(n)}}(z,a)$ and $F_T(z,a) = C_T^{\bnu}(z,a)$. Since $c$ is continuous and $\nu_{T,1}^{(n)}$ weakly converges to $\nu_{T,1}$, we have that $J^{\bnu^{(n)}}_{*,T}$ continuously converges to $J^{\bnu}_{*,T}$ by \citet[Proposition 7.32]{BeSh78}. But this implies that $F_{T-1}^{(n)}$ continuously converges to $F_{T-1}$, and so, $J^{\bnu^{(n)}}_{*,T-1}$ continuously converges to $J^{\bnu}_{*,T-1}$ again by \citet[Proposition 7.32]{BeSh78}. Then, by the induction hypothesis, we can conclude that $F_t^{(n)}$ and $J_{*,t}^{\bnu^{(n)}}$ continuously converge to $F_t$ and $J^{\bnu}_{*,t}$, respectively, for each $t\leq T$. Therefore, Theorems~\ref{thm:MFE} and \ref{appr-thm} hold for the finite-horizon cost criterion under the same assumptions. Furthermore, if we start the mean-field game at time $\tau>0$ with initial measure $\mu_{\tau}$, then the pair $\bigl(\{\pi_t\}_{\tau\leq t \leq T},\{\mu_t\}_{\tau\leq t \leq T}\bigr)$ in Theorem~\ref{thm:MFE} is still a mean-field equilibrium for the sub-game.\Halmos
\end{remark}

\section{An Example}\label{example}

In this section, we consider a specific additive noise model to illustrate our results. In this model, the state and observation dynamics of a generic agent for the mean-field game are given respectively by
\begin{align}
x(t+1) &= \int_{\sX} f(x(t),a(t),x)  \mu_t(dx) + g(x(t),a(t))  w(t) \nonumber \\
&\eqqcolon F(x(t),a(t),\mu_t) + g(x(t),a(t)) w(t) \nonumber \\
\intertext{and}
y(t) &= \int_{\sX} h(x(t),x)  \mu_t(dx) +  v(t)  \nonumber \\
&\eqqcolon H(x(t),\mu_t) +  v(t), \nonumber
\end{align}
where $x(t) \in \sX$, $y(t) \in \sY$, $a(t) \in \sA$, $w(t) \in \sW$, and $v(t) \in \sV$. Here, we assume that $\sX = \sY = \sW =\sV = \R$, $\sA \subset \R$, and $\{w(t)\}$ and $\{v(t)\}$  are sequences of i.i.d. standard normal random variables independent of each other. The one-stage cost function of a generic agent is given by
\begin{align}
c(x(t),a(t),\mu_t) = \int_{\sX} d(x(t),a(t),x) \phantom{i} \mu_t(dx), \nonumber
\end{align}
for some measurable function $d: \sX \times \sA \times \sX \rightarrow [0,\infty)$.

This model is the infinite-population limit of the $N$-agent game model with state and observation dynamics
\begin{align}
x_i^N(t+1) &= \frac{1}{N} \sum_{j=1}^N f(x_i^N(t),a_i^N(t),x_j^N(t)) + g(x_i^N(t),a_i^N(t)) w_i^N(t) \nonumber \\ y_i^N(t) &= \frac{1}{N} \sum_{j=1}^N h(x_i^N(t),x_j^N(t)) + v_i^N(t) \nonumber
\end{align}
and the one-stage cost function
\begin{align}
c(x_i^N(t),a_i^N(t),e_t^{(N)}) &= \frac{1}{N} \sum_{j=1}^N d(x_i^N(t),a_i^N(t),x_j^N(t)). \nonumber
\end{align}

For this model, Assumption~\ref{as1} holds with $w(x) = 1 + x^2$ and $\alpha = \max\{1 + \|f\|^2, L\}$ under the following conditions: (i) $\sA$ is compact, (ii) $d$ is continuous and bounded, (iii) $g$ is continuous, and $f$ is bounded and continuous, (iv) $\sup_{a \in \sA} g^2(x,a) \leq L x^2$ for some $L>0$, (v) $h$ is continuous and bounded. Note that $\|f\|$ is defined as
\begin{align}
\|f\| \coloneqq \sup_{(x,a,x') \in \sX \times \sA \times \sX} |f(x,a,x')|. \nonumber
\end{align}

Indeed, we have
\begin{align}
\int_{\sX} (1 + y^2) p(dy|x,a,\mu)  &= \int_{\sX} \biggl( 1+ \bigl[ F(x,a,\mu) + g(x,a) y \bigr]^2 \biggr)  q(y) m(dy) \nonumber \\
&\leq 1 + \|f\|^2 + g^2(x,a) \leq \alpha (1+x^2), \nonumber
\end{align}
where $q$ is the standard normal density and $m$ is the Lebesgue measure. Hence, Assumption~\ref{as1}-(e) holds. In order to verify Assumption~\ref{as1}-(b), suppose $(x_n,a_n,\mu_n) \rightarrow (x,a,\mu)$ and let $g \in C_b(\sX)$. Then, we have
\begin{align}
\lim_{n\rightarrow\infty} \int_{\sX} g(y) p(dy|x_n,a_n,\mu_n)
&= \lim_{n\rightarrow\infty} \int_{\sX} g\bigl(F(x_n,a_n,\mu_n) + g(x_n,a_n) z\bigl) q(z) m(dz) \nonumber \\
&= \int_{\sX} g\bigl(F(x,a,\mu) + g(x,a) z\bigl) q(z) m(dz) \nonumber
\end{align}
since $g$ and $F$ are continuous, where the continuity of $F$ follows from \citet[Theorem 3.5]{Lan81} and the fact that $f$ is bounded and continuous. Therefore, the transition probability $p(\,\cdot\,|x,a,\mu)$ is weakly continuous. Thus, Assumption~\ref{as1}-(b) holds. Note that Assumption~\ref{as1}-(f) holds if the initial distribution $\mu_0$ has a finite second moment. Assumption~\ref{as1}-(a) trivially holds since $d$ is bounded and continuous. Finally, we will verify Assumption~\ref{as1}-(c). Suppose $(x_n,\mu_n) \rightarrow (x,\mu)$. Then, we have $r(dy|x_n,\mu_n) = q\bigl(y-H(x_n,\mu_n)\bigr) m(dy)$ and $r(dy|x,\mu) = q\bigl(y-H(x,\mu)\bigr) m(dy)$. Since $q\bigl(y-H(x_n,\mu_n)\bigr) \rightarrow q\bigl(y-H(x,\mu)\bigr)$ as $n\rightarrow\infty$ for all $y \in \sY$, by Scheff\'{e}'s theorem (see, e.g., \citet[Theorem 16.12]{Bil95}) we have $r(\,\cdot\,|x_n,\mu_n) \rightarrow r(\,\cdot\,|x_n,\mu_n)$ as $n\rightarrow\infty$ in total variation norm. Thus, Assumption~\ref{as1}-(c) holds. Therefore, under (i)-(v), there exists a mean-field equilibrium for the mean-field game of this example.

For the same model, Assumption~2-(a),(c) holds under the following conditions: (vi) $d(x,a,y)$ is (uniformly) Lipschitz in $y$ with Lipschitz constant $K_d$, (vii) $f(x,a,y)$ is (uniformly) Lipschitz in $y$ with Lipschitz constant $K_f$, (viii) $g$ is bounded and $\inf_{(x,a) \in \sX \times \sA} |g(x,a)| \eqqcolon \theta > 0$, and (ix) $H$ is only a function of $x$.

Indeed, we have
\begin{align}
\omega_c(r) &= \sup_{(x,a) \in \sX\times\sA} \sup_{\substack{\mu,\nu: \\ d_{BL}(\mu,\nu)\leq r}} \biggl |\int_{\sX} d(x,a,y) \mu(dy) - \int_{\sX} d(x,a,y) \nu(dy) \biggr| \nonumber \\
&\leq \sup_{(x,a) \in \sX\times\sA} \sup_{\substack{\mu,\nu: \\ d_{BL}(\mu,\nu)\leq r}} L_d d_{BL}(\mu,\nu) \nonumber \\
&\leq L_d r, \nonumber
\end{align}
where $L_d \coloneqq \max\{\|d\|,K_d\}$. Hence, $\omega_c(r) \rightarrow 0$ as $r\rightarrow0$. For $\omega_p$, we have
\begin{align}
\omega_{p}(r) &= \sup_{(x,a) \in \sX\times\sA} \sup_{\substack{\mu,\nu: \\ d_{BL}(\mu,\nu)\leq r}} \|p(\,\cdot\,|x,a,\mu) - p(\,\cdot\,|x,a,\nu)\|_{TV} \nonumber \\
&= \sup_{(x,a) \in \sX\times\sA} \sup_{\substack{\mu,\nu: \\ d_{BL}(\mu,\nu)\leq r}}
\sup_{\|l\| \leq 1} \biggl| \int_{\sX} l\bigl(F(x,a,\mu)+g(x,a) z\bigr) q(z) m(dz) \nonumber \\
&\phantom{xxxxxxxxxxxxxxxxxxxxxxxxx}-  \int_{\sX} l(F(x,a,\nu)+g(x,a) z) q(z) m(dz) \biggr| \nonumber \\
&= \sup_{(x,a) \in \sX\times\sA} \sup_{\substack{\mu,\nu: \\ d_{BL}(\mu,\nu)\leq r}}
\sup_{\|l\| \leq 1} \biggl| \int_{\sX} l\bigl(g(x,a) z\bigr) q\biggl(z-\frac{F(x,a,\mu)}{g(x,a)}\biggr) m(dz) \nonumber \\
&\phantom{xxxxxxxxxxxxxxxxxxxxxxxxx}-  \int_{\sX} l\bigl(g(x,a) z\bigr) q\biggl(z-\frac{F(x,a,\nu)}{g(x,a)}\biggr) m(dz) \biggr| \nonumber \\
&\leq \sup_{(x,a) \in \sX\times\sA} \sup_{\substack{\mu,\nu: \\ d_{BL}(\mu,\nu)\leq r}}
\sup_{\|h\|\leq 1} \biggl| \int_{\sX} h(z) q\biggl(z-\frac{F(x,a,\mu)}{g(x,a)}\biggr) m(dz) \nonumber \\
&\phantom{xxxxxxxxxxxxxxxxxxxxxxxxx}- \int_{\sX} h(z) q\biggl(z-\frac{F(x,a,\nu)}{g(x,a)}\biggr) m(dz) \biggr|. \label{bound}
\end{align}
For any compact interval $K = [-k,k] \subset \sX$, we can upper bound (\ref{bound}) as follows:
\begin{align}
&(\ref{bound}) \leq \sup_{(x,a) \in \sX\times\sA} \sup_{\substack{\mu,\nu: \\ d_{BL}(\mu,\nu)\leq r}}
\sup_{\|h\|\leq 1} \biggl( \biggl| \int_{K} h(z) q\biggl(z-\frac{F(x,a,\mu)}{g(x,a)}\biggr) m(dz) - \int_{K} h(z) q\biggl(z-\frac{F(x,a,\nu)}{g(x,a)}\biggr) m(dz) \biggr| \nonumber \\
&\phantom{xxxxxxxxxxx}+ \int_{K^c} q\biggl(z-\frac{F(x,a,\mu)}{g(x,a)}\biggr) m(dz) + \int_{K^c}  q\biggl(z-\frac{F(x,a,\nu)}{g(x,a)}\biggr) m(dz) \biggr). \label{bound2}
\end{align}
The last two integrals in the last expression go to zero (uniformly in $(x,a,\mu,\nu)$) as $k \rightarrow \infty$, since $F$ and $g$ are bounded, and $\inf_{(x,a) \in \sX \times \sA} |g(x,a)| > 0$. For any $\varepsilon>0$, let $K_{\varepsilon}=[-k_{\varepsilon},k_{\varepsilon}] \subset \sX$ so that the sum of these integrals is less than $\varepsilon$ for all $(x,a,\mu,\nu)$. Let $T_{\varepsilon}$ denote the Lipschitz seminorm of $q$ on $K_{\varepsilon}$. Then, we have
\begin{align}
(\ref{bound2}) &\leq  \sup_{(x,a) \in \sX\times\sA} \sup_{\substack{\mu,\nu: \\ d_{BL}(\mu,\nu)\leq r}}
\sup_{\|h\|\leq 1} \biggl| \int_{K_{\varepsilon}} h(z) q\biggl(z-\frac{F(x,a,\mu)}{g(x,a)}\biggr) m(dz) - \int_{K_{\varepsilon}} h(z) q\biggl(z-\frac{F(x,a,\nu)}{g(x,a)}\biggr) m(dz) \biggr| + \varepsilon \nonumber \\
&\leq T_{\varepsilon} \sup_{(x,a) \in \sX\times\sA} \sup_{\substack{\mu,\nu: \\ d_{BL}(\mu,\nu)\leq r}}
\sup_{\|h\|\leq 1} \int_{K_{\varepsilon}} |h(z)|  \biggl|\frac{F(x,a,\mu)}{g(x,a)} - \frac{F(x,a,\nu)}{g(x,a)} \biggr| m(dz) + \varepsilon \nonumber \\
&\leq \frac{T_{\varepsilon}}{\theta} m(K_{\varepsilon})  \sup_{(x,a) \in \sX\times\sA} \sup_{\substack{\mu,\nu: \\ d_{BL}(\mu,\nu)\leq r}} |F(x,a,\mu) - F(x,a,\nu)| + \varepsilon \nonumber \\
&\leq \frac{T_{\varepsilon}}{\theta} m(K_{\varepsilon}) \sup_{(x,a) \in \sX\times\sA} \sup_{\substack{\mu,\nu: \\ d_{BL}(\mu,\nu)\leq r}} L_f d_{BL}(\mu,\nu) + \varepsilon \nonumber \\
&\leq \frac{T_{\varepsilon}}{\theta} m(K_{\varepsilon}) L_f r + \varepsilon, \nonumber
\end{align}
where $L_f \coloneqq \max\{\|f\|,K_f\}$. Since $\varepsilon$ is arbitrary, we have $\omega_p(r) \rightarrow 0$ as $r\rightarrow0$. Thus, Assumption~\ref{as2}-(a) holds. Note that Assumption~\ref{as2}-(c) automatically holds as $H$ is only a function of $x$.


To establish Assumption~\ref{as2}-(b), we can impose the following additional assumption, as we did in Remark~\ref{continuity}. Suppose $\bmu$ is the measure-flow in mean-field equilibrium.
\begin{itemize}
\item [(b')] For $\bmu \in \Xi$, there a exists unique minimizer $a_z \in \sA$ of
\begin{align}
C_t^{\bnu}(z,\,\cdot\,) + \beta \int_{\sZ} J_{*,t+1}^{\bmu}(z') \eta_t^{\bnu}(dz'|z,\,\cdot\,), \nonumber
\end{align}
for each $z \in \sZ$ and for all $t\geq0$.
\end{itemize}
Under assumption (b'), one can prove that Assumption~\ref{as2}-(b) holds. Note that, by Remark~\ref{continuity}, assumption (b') is true if, for instance, $d(x,a,y)$ and $\varrho(x'|x,a,\mu)$ are strictly convex in $a$, where $\varrho$ is given by
\begin{align}
\varrho(x'|x,a,\mu) = q\biggl(\frac{x'-F(x,a,\mu)}{g(x,a)}\biggr) \frac{1}{g(x,a)}. \nonumber
\end{align}


\section{Conclusion}\label{conc}

This paper has considered discrete-time partially observed mean-field games subject to infinite-horizon discounted cost, for Polish state, observation, and action spaces. Under mild conditions, the existence of a Nash equilibrium has been established for this game model using the conversion of partially observed Markov decision processes to fully observed Markov decision processes in the belief space and then using the dynamic programming principle. We have also established that the mean-field equilibrium policy, when used by each agent, constitutes a nearly Nash equilibrium for games with sufficiently many agents.

One interesting future direction of research to pursue is to study partially observed \emph{team} problems of the mean-field type. In this case, one possible approach is to establish the global optimality of person-by-person optimal policies under some convexity assumptions and then use the results developed in this paper. Finally, partially observed mean-field games with average-cost and risk-sensitive optimality criteria are also worth studying. In particular, using the vanishing discount factor approach in MDP theory (i.e., with discount factor $\beta \rightarrow 1$), it might be possible to establish similar results for the average cost case.

\section*{Appendix}

\appsec

\subsection{Proof of Proposition~\ref{prop:belief_conv}}\label{app1}

Fix any $(z,a) \in \sZ \times \sA$ and $t\geq0$. To ease the notation, let us write $\bmu^{(n)} \coloneqq \bmu^{\bnu^{(n)}}$ and $\bmu \coloneqq \bmu^{\bnu}$. First note that, for all $t\geq0$, $\mu_t^{(n)}$ weakly converges to $\mu_t$ as $n\rightarrow\infty$.

We will mimic the proof technique used in \citet[Section 5]{FeKaZg16} to prove the result. To this end, we first prove the following lemma.

\begin{lemma}\label{lemma:aux}
Fix any $(z,a) \in \sZ \times \sA$. Then, for any $f \in C_b(\sX)$, we have
\begin{align}
\lim_{n\rightarrow\infty} \sup_{C \in \B(\sY)} \biggl| \int_{C} F_t^{(n)}(z,a,y)(f) \text{ } H_t^{(n)}(dy|z,a) - \int_{C} F_t(z,a,y)(f) \text{ } H_t(dy|z,a) \biggr|=0. \nonumber
\end{align}
In particular, if $f \equiv 1$, then the above result implies that $H_t^{(n)}(\,\cdot\,|z,a)$ converges to $H_t(\,\cdot\,|z,a)$ in total variation norm.
\end{lemma}

\smallskip

\proof{Proof.}
We have
\begin{align}
&\sup_{C \in \B(\sY)} \biggl| \int_{C} F_t^{(n)}(z,a,y)(f) \text{ } H_t^{(n)}(dy|z,a) - \int_{C} F_t(z,a,y)(f) \text{ } H_t(dy|z,a) \biggr| \nonumber \\
&= \sup_{C \in \B(\sY)} \biggl| \int f(x) 1_{C}(y) r(dy|x,\mu_{t+1}^{(n)}) p(dx|x',a,\mu_t^{(n)}) z(dx') \nonumber \\
&\phantom{xxxxxxxxxxxxxxxxxxxxxx}- \int f(x) 1_{C}(y) r(dy|x,\mu_{t+1}) p(dx|x',a,\mu_t) z(dx') \biggr| \nonumber \\
&\leq \sup_{C \in \B(\sY)} \biggl| \int f(x) 1_{C}(y) r(dy|x,\mu_{t+1}^{(n)}) p(dx|x',a,\mu_t^{(n)}) z(dx') \nonumber \\
&\phantom{xxxxxxxxxxxxxxxxxxxxxx}- \int f(x) 1_{C}(y) r(dy|x,\mu_{t+1}) p(dx|x',a,\mu_t^{(n)}) z(dx') \biggr| \nonumber \\
&+\sup_{C \in \B(\sY)} \biggl| \int f(x) 1_{C}(y) r(dy|x,\mu_{t+1}) p(dx|x',a,\mu_t^{(n)}) z(dx') \nonumber \\
&\phantom{xxxxxxxxxxxxxxxxxxxxxx}- \int f(x) 1_{C}(y) r(dy|x,\mu_{t+1}) p(dx|x',a,\mu_t) z(dx') \biggr| \nonumber \\
&\leq  \int \|f\| \|r(\,\cdot\,|x,\mu_{t+1}^{(n)})-r(\,\cdot\,|x,\mu_{t+1})\|_{TV} p(dx|x',a,\mu_t^{(n)}) z(dx') \nonumber \\
&\phantom{xxxxx}+\int \sup_{C \in \B(\sY)} \biggl| \int_{\sX} f(x) r(C|x,\mu_t{+1}) p(dx|x',a,\mu_t^{(n)}) - \int f(x) r(C|x,\mu_t{+1}) p(dx|x',a,\mu_t) \biggr| z(dx'). \nonumber
\end{align}
The first term in the last expression converges to zero as $n\rightarrow\infty$ by \citet[Theorem 3.5]{Lan81} since $p(\,\cdot\,|x',a,\mu_t^{(n)})$ weakly converges to $p(dx|x',a,\mu_t)$ and $\|r(\,\cdot\,|x,\mu_{t+1}^{(n)})-r(\,\cdot\,|x,\mu_{t+1})\|$ converges continuously to $0$. For the second term, define ${\cal F} \coloneqq \{f(\,\cdot\,) r(C|\,\cdot\,,\mu_{t+1}): C \in \B(\sY)\}$. Observe that ${\cal F}$ is an equicontinuous family of functions. Indeed, let $x_n \rightarrow x$ in $\sX$. Then, we have
\begin{align}
&\lim_{n\rightarrow\infty} \sup_{C \in \B(\sY)} |f(x_n)r(C|x_n,\mu_{t+1})-f(x)r(C|x,\mu_{t+1})| \nonumber \\
&\leq \lim_{n\rightarrow\infty} \biggl( \sup_{C \in \B(\sY)} |f(x_n)r(C|x_n,\mu_{t+1})-f(x_n)r(C|x,\mu_{t+1})| \nonumber \\
&\phantom{xxxxxxxxxxxxxxxxxxxx}+ \sup_{C \in \B(\sY)} |f(x_n)r(C|x,\mu_{t+1})-f(x)r(C|x,\mu_{t+1})| \biggr) \nonumber \\
&\leq \lim_{n\rightarrow\infty} \bigl( |f(x_n)| \|r(\,\cdot\,|x_n,\mu_{t+1}) - r(\,\cdot\,|x,\mu_{t+1})\|_{TV} + |f(x_n) - f(x)| \bigr) =0. \nonumber
\end{align}
Since ${\cal F}$ is also uniformly bounded, the second term in the last expression also goes to zero as $n\rightarrow\infty$ since $p(\,\cdot\,|x',a,\mu_t^{(n)})$ weakly converges to $p(\,\cdot\,|x',a,\mu_t)$.\Halmos
\endproof

Let $\{f_k\} \subset C_b(\sX)$ be the weak convergence determining class of functions in $C_b(\sX)$; that is, $\mu_n$ weakly converges to $\mu$ in $\P(\sX)$ if and only if $\lim_{n\rightarrow\infty}\mu_n(f_k)=\mu(f_k)$ for all $k$.

Now, we prove that for any subsequence $\{F_t^{(n_l)}(z,a,y)\}_{l\geq1}$ of $\{F_t^{(n)}(z,a,y)\}_{n\geq1}$, there exists a further subsequence $\{F_t^{(n_{l_m})}(z,a,y)\}_{m\geq1}$ such that $F_t^{(n_{l_m})}(z,a,y)$ weakly converges to $F_t(z,a,y)$ for $H_t(\,\cdot\,|z,a)$-almost everywhere. Let us write the subsequence $\{F_t^{(n_l)}(z,a,y)\}_{l\geq1}$ as $\{F_t^{(l,0)}(z,a,y)\}_{l\geq1}$. Since, by Lemma~\ref{lemma:aux}
\begin{align}
&\lim_{l\rightarrow\infty} \sup_{C \in \B(\sY)} \biggl| \int_{C} F_t^{(l,0)}(z,a,y)(f_1) \text{ } H_t^{(l,0)}(dy|z,a) - \int_{C} F_t(z,a,y)(f_1) \text{ } H_t(dy|z,a) \biggr|=0, \nonumber
\end{align}
$F_t^{(l,0)}(z,a,y)(f_1)$ converges in probability $H_t(\,\cdot\,|z,a)$ to $F_t(z,a,y)(f_1)$ by \citet[Theorem 5.2]{FeKaZg16}, and so, there is a subsequence $\{F_t^{(l,1)}(z,a,y)(f_1)\}$ of $\{F_t^{(l,0)}(z,a,y)(f_1)\}$ such that $F_t^{(l,1)}(z,a,y)(f_1)$ converges to $F_t(z,a,y)(f_1)$ $H_t(\,\cdot\,|z,a)$-almost everywhere. Similarly, by Lemma~\ref{lemma:aux}
\begin{align}
&\lim_{l\rightarrow\infty} \sup_{C \in \B(\sY)} \biggl| \int_{C} F_t^{(l,1)}(z,a,y)(f_2) \text{ } H_t^{(l,1)}(dy|z,a) - \int_{C} F_t(z,a,y)(f_2) H_t(dy|z,a) \biggr|=0, \nonumber
\end{align}
and so, $F_t^{(l,1)}(z,a,y)(f_2)$ converges in $H_t(\,\cdot\,|z,a)$-probability to $F_t(z,a,y)(f_2)$ by \citet[Theorem 5.2]{FeKaZg16}. Therefore, there is a subsequence $\{F_t^{(l,2)}(z,a,y)(f_2)\}$ of $\{F_t^{(l,1)}(z,a,y)(f_2)\}$ such that $F_t^{(l,2)}(z,a,y)(f_2)$ converges to $F_t(z,a,y)(f_2)$ $H_t(\,\cdot\,|z,a)$-almost everywhere. Continuing in this manner, we obtain an array of sequences. Then, by Cantor's diagonal argument, for all $k\geq1$, $F_t^{(m,m)}(z,a,y)(f_k)$ converges to $F_t(z,a,y)(f_k)$ $H_t(\,\cdot\,|z,a)$-almost everywhere as $m\rightarrow\infty$. This implies that $F_t^{(m,m)}(z,a,y)$ weakly converges to $F_t(z,a,y)$ $H_t(\,\cdot\,|z,a)$-almost everywhere.

Now, we combine this result and convergence of $H_t^{(n)}(\,\cdot\,|z,a)$ to $H_t(\,\cdot\,|z,a)$ in total variation norm to complete the proof. By the portmanteau theorem, it is sufficient to prove that $\liminf_{n\rightarrow\infty} \eta_t^{(n)}(D|z,a) \geq \eta_{t}(D|z,a)$ for all $D$ open in $\sZ$. Suppose to the contrary that there exists an open set $D \subset \sZ$ such that
$\liminf_{n\rightarrow\infty} \eta_t^{(n)}(D|z,a) < \eta_{t}(D|z,a)$. Then, there exists a subsequence $\{\eta_t^{(n_k)}(D|z,a)\}$ of $\{\eta_t^{(n)}(D|z,a)\}$ such that $\eta_t^{(n_k)}(D|z,a) \leq \eta_{t}(D|z,a)-\varepsilon$ for all $k$. By the above, there exists a subsequence $\{F_t^{(n_{k_l})}(D|z,a)\}$ of $\{F_t^{(n_k)}(D|z,a)\}$ such that $F_t^{(n_{k_l})}(z,a,y)$ weakly converges to $F_t(z,a,y)$ $H_t(\,\cdot\,|z,a)$-almost everywhere. Since $D$ is open, we have
\begin{align}
\liminf_{l\rightarrow\infty} 1_{\bigl\{F_t^{(n_{k_l})}(z,a,y) \in D\bigr\}} \geq 1_{\bigl\{F_t(z,a,y) \in D\bigr\}}, \text{ } \text{$H_t(\,\cdot\,|z,a)$-a.e.} \nonumber
\end{align}
Then by \citet[Lemma 5.1(i)]{FeKaZg16} and the fact that $H_t^{(n_{k_l})}(\,\cdot\,|z,a)$ converges to $H_t(\,\cdot\,|z,a)$ in total variation norm, we have
\begin{align}
\liminf_{l\rightarrow\infty} \eta_{t}^{(n_{k_l})}(D|z,a)  &= \liminf_{l\rightarrow\infty} \int_{\sY} 1_{\bigl\{F_t^{(n_{k_l})}(z,a,y) \in D\bigr\}} H_t^{(n_{k_l})}(dy|z,a) \nonumber \\
&\geq  \int_{\sY} \liminf_{l\rightarrow\infty} 1_{\bigl\{F_t^{(n_{k_l})}(z,a,y) \in D\bigr\}} H_t(dy|z,a) \nonumber \\
&\geq  \int_{\sY} 1_{\bigl\{F_t(z,a,y) \in D\bigr\}} H_t(dy|z,a) = \eta_t(D|z,a), \nonumber
\end{align}
which is a contradiction. Hence, we must have $\liminf_{n\rightarrow\infty} \eta_t^{(n)}(D|z,a) \geq \eta_{t}(D|z,a)$ for all $D$ open in $\sZ$.

\subsection{Proof of Proposition~\ref{app-prop1}}\label{app3}

Fix any $N\geq1$ and ${\boldsymbol \pi}^{(N)} \in {\bf \Pi}^{(N)}$. For each $t\geq0$, let $\hat{\rP}_t$
denote the probability law of the states $\bs^N(t) = (s^N_1(t),\ldots,s^N_N(t))$ and the actions $\ba^N(t) = (a^N_1(t),\ldots,a^N_N(t))$ under ${\boldsymbol \pi}^{(N)}$ in the new game model. Similarly, let $\rP_t$
denote the probability law of the states $\bx^N(t) = (x^N_1(t),\ldots,x^N_N(t))$, the observations $\by^N(k) = (y^N_1(k),\ldots,y^N_N(k))$ for $k=0,\ldots,t$, and the actions $\ba^N(t) = (a^N_1(t),\ldots,a^N_N(t))$ under ${\boldsymbol \pi}^{(N)}$ in the original finite agent game model. We prove that, for each $t\geq0$,
\begin{align}
\hat{\rP}_t = \rP_t, \nonumber
\end{align}
which implies that $\hat{J}_i^N({\boldsymbol \pi}^{(N)}) = J_i^N({\boldsymbol \pi}^{(N)})$ for all $i=1,\ldots,N$.

The claim trivially holds for $t=0$. Suppose that the claim holds for $t$ and consider $t+1$. For $t+1$, let
\begin{align}
{\boldsymbol A}_{t+1} &= A_{t+1}^1 \times \ldots \times A_{t+1}^N, \text{ } \text{$A_{t+1}^i \in \B(\sX)$ for $i=1,\ldots,N$}, \nonumber \\
{\boldsymbol B}_{k} &= B_{k}^1 \times \ldots \times B_{k}^N, \text{ } \text{$B_{k}^i \in \B(\sY)$ for $i=1,\ldots,N$} \text{and $k=0,\ldots,t+1$}, \nonumber \\
{\boldsymbol D}_{t+1} &= D_{t+1}^1 \times \ldots \times D_{t+1}^N, \text{ } \text{$D_{t+1}^i \in \B(\sA)$ for $i=1,\ldots,N$}. \nonumber
\end{align}
Define the following transition probability $Q_t$ on $\prod_{i=1}^{N} \sS_{t+1}$ given $\prod_{i=1}^{N} \sS_{t} \times \prod_{i=1}^N \sA$ as:
\begin{align}
Q_t(d\bs'|\bs,\ba) \coloneqq \prod_{i=1}^N P_t(ds'_i|s_i,a_i,F_N(\bs)), \nonumber
\end{align}
where $F_N(\bs) \coloneqq \frac{1}{N} \sum_{i=1}^N \delta_{s_i}$. Define also $G_{t+1} \coloneqq {\boldsymbol A}_{t+1} \times {\boldsymbol B}_{t+1} \times \ldots {\boldsymbol B}_{0} \times {\boldsymbol D}_{t+1}$, $L_t \coloneqq {\boldsymbol B}_{t} \times \ldots {\boldsymbol B}_{0}$, and $U_{t+1} \coloneqq {\boldsymbol A}_{t+1} \times {\boldsymbol B}_{t+1} \times {\boldsymbol D}_{t+1}$. Then, we have
\begin{align}
&\hat{\rP}_t(G_{t+1}) \nonumber \\
&= \int_{(\sS_t)^t \times \sA^t} \int_{G_{t+1}} \biggl( \prod_{i=1}^N \pi_{t+1}^i(da^N_i(t+1)|s^N_i(t+1)) \biggr) Q_t(d\bs^N(t+1)|\bs^N(t),\ba^N(t)) \hat{\rP}_t(d\bs^N(t),d\ba^N(t))  \nonumber \\
&= \int_{(\sS_t)^t \times \sA^t} \int_{G_{t+1}} \prod_{i=1}^N \biggl( \pi_{t+1}^i(da^N_i(t+1)|s^N_i(t+1))  \nonumber \\
&\phantom{xxxxxxxxxxxxxxxxxxxx} r(dy^N_i(t+1)|x^N_i(t+1)) p(dx^N_i(t+1)|x^N_i(t),a^N_i(t),e_t^{(N)}) \nonumber \\
&\phantom{xxxxxxxxxxxxxxxxxxxxxxxxxxxxxxxx}\prod_{k=0}^t \delta_{y^N_i(k)}(dy^N_i(k)) \biggr) \rP_t(d\bx^N(t),d\by^N(t:0),d\ba^N(t))  \nonumber \\
&= \int_{\sX^t \times \sA^t \times L_t} \int_{U_{t+1}} \prod_{i=1}^N \biggl( \pi_{t+1}^i(da^N_i(t+1)|s^N_i(t+1))  \nonumber \\
&\phantom{xxxxxxxxxxxxxxxxxxx} r(dy^N_i(t+1)|x^N_i(t+1)) p(dx^N_i(t+1)|x^N_i(t),a^N_i(t),e_t^{(N)})  \biggr) \nonumber \\
&\phantom{xxxxxxxxxxxxxxxxxxxxxxxxxxxxxxxxxxxxxxxxxxxxxxxxx} \rP_t(d\bx^N(t),d\by^N(t:0),d\ba^N(t))  \nonumber \\
&= \rP_{t+1}(G_{t+1}), \nonumber
\end{align}
where $d\by^N(k:0) \coloneqq (d\by^N(k),\ldots,d\by^N(0))$. Hence, $\hat{\rP}_{t+1} = \rP_{t+1}$. This implies that $\hat{J}_i^N({\boldsymbol \pi}^{(N)}) = J_i^N({\boldsymbol \pi}^{(N)})$ for all $i=1,\ldots,N$.

The second part of the proposition can be proved similarly, so we omit the details.

\subsection{Proof of Lemma~\ref{app-lemma1}}\label{app4}

The claim trivially holds for $t=0$. Suppose that the claim holds for $t$ and consider $t+1$. For $t+1$, let $G_{t+1} = A_{t+1} \times B_{t+1} \times \ldots B_{0}$, where $A_{t+1} \in \B(\sX)$ and $B_k \in \B(\sY)$ for $k=0,\ldots,t+1$. Define also $L_t \coloneqq B_{t} \times \ldots B_{0}$, and $U_{t+1} \coloneqq A_{t+1} \times B_{t+1}$. Then, we have
\begin{align}
&\int_{\sS \times \sA} P_t(G_{t+1}|s(t),a(t),\Delta_t) \pi_t(da(t)|s(t)) \Delta_t(ds(t)) \nonumber \\
&= \int_{\sS \times \sA} \int_{G_{t+1}} r(dy(t+1)|x(t+1)) p(dx(t+1)|x(t),a(t),\Delta_{t,1}) \prod_{k=0}^t \delta_{y(k)}(dy(k)) \pi_t(da(t)|s(t)) \Delta_t(ds(t)) \nonumber \\
&= \int_{\sX \times \sA \times L_t} \int_{U_{t+1}} r(dy(t+1)|x(t+1)) p(dx(t+1)|x(t),a(t),\Delta_{t,1}) \pi_t(da(t)|s(t)) \Delta_t(ds(t)) \nonumber \\
&= \Delta_t(G_{t+1}). \nonumber
\end{align}
Since, $G_{t+1}$ is arbitrary, this completes the proof.

\subsection{Proof of Proposition~\ref{prop9}}\label{app6}

Fix any sequence $\{T_N\}_{N\geq1}$ satisfying the hypothesis of the proposition. Fix any $t\geq0$ and suppose that
\begin{align}
\lim_{N\rightarrow\infty}\bigl| {\cal L}(\ts_1^N(t))(g_N) - {\cal L}(\hs^N(t))(g_N) \bigr| = 0 \label{neweq1}
\end{align}
for any bounded sequence $\{g_N\}_{N\geq1} \subset C_b(\sS_t)$. Given (\ref{neweq1}), we prove that
\begin{align}
\lim_{N\rightarrow\infty} \bigl| {\cal L}(\ts_1^N(t),\tilde{\Delta}_t^{(N)})(T_N) - {\cal L}(\hs^N(t),\delta_{\Delta_t})(T_N) \bigr| = 0. \label{neweq2}
\end{align}
Indeed, we have
\begin{align}
&\bigl| {\cal L}(\ts_1^N(t),\tilde{\Delta}_t^{(N)})(T_N) - {\cal L}(\hs^N(t),\delta_{\Delta_t})(T_N) \bigr| \nonumber \\
&\phantom{xxx}\leq \biggl| \int_{\sS_t \times \P(\sS_t)} T_N(s,\Delta) {\cal L}(\ts_1^N(t),\tilde{\Delta}_t^{(N)})(ds,d\Delta) - \int_{\sS_t \times \P(\sS_t)} T_N(s,\Delta) {\cal L}(\ts_1^N(t),\delta_{\Delta_t})(ds,d\Delta) \biggr| \nonumber \\
&\phantom{xxx}+ \biggl| \int_{\sS_t \times \P(\sS_t)} T_N(s,\Delta) {\cal L}(\ts_1^N(t),\delta_{\Delta_t})(ds,d\Delta) - \int_{\sS_t \times \P(\sS_t)} T_N(s,\Delta) {\cal L}(\hs^N(t),\delta_{\Delta_t})(ds,d\Delta) \biggr|. \label{eq12}
\end{align}
First, note that since $\{T_N(\,\cdot\,,\Delta_t)\}_{N\geq1} \subset C_b(\sS_t)$, we have
\begin{align}
\lim_{N\rightarrow\infty} \biggl| \int_{\sS_t} T_N(s,\Delta_t) {\cal L}(\ts_1^N(t))(ds) - \int_{\sS_t} T_N(s,\Delta_t) {\cal L}(\hs^N(t))(ds) \biggr|=0 \nonumber
\end{align}
by (\ref{neweq1}).

Now, let us consider the first term in (\ref{eq12}). To that end, define ${\cal F} \coloneqq \bigl\{T_N(s,\,\cdot\,): s \in \sS_t, N\geq1)\bigr\}$. Note that ${\cal F}$ is a uniformly bounded and equicontinuous family of functions on $\P(\sS_t)$, and therefore
\begin{align}
\lim_{N\rightarrow\infty} E\biggl[ \sup_{F \in {\cal F}} \bigl| F(\tilde{\Delta}_t^{(N)}) - F(\Delta_t) \bigr| \biggr] = 0 \nonumber
\end{align}
as ${\cal L}(\tilde{\Delta}_t^{(N)}) \rightarrow {\cal L}(\Delta_t)$ weakly. Then, we have
\begin{align}
&\lim_{N\rightarrow\infty}\biggl| \int_{\sS_t \times \P(\sS_t)} T_N(s,\Delta) {\cal L}(\ts_1^N(t),\tilde{\Delta}_t^{(N)})(ds,d\Delta) - \int_{\sS_t \times \P(\sS_t)} T_N(s,\Delta) {\cal L}(\ts_1^N(t),\delta_{\Delta_t})(ds,d\Delta) \biggr| \nonumber \\
&\leq \lim_{N\rightarrow\infty} \int_{\sS_t} \biggl| \int_{\P(\sS_t)} T_N(s,\Delta) {\cal L}(\tilde{\Delta}_t^{(N)}|\ts_1^N(t))(d\Delta|s) - \int_{\P(\sS_t)} T_N(s,\Delta) {\cal L}(\delta_{\Delta_t})(d\Delta) \biggr| {\cal L}(\ts_1^N(t))(ds) \nonumber \\
&\leq \lim_{N\rightarrow\infty} E\biggl[ E\biggl[ \bigl|T_N(\ts_1^N(t),\tilde{\Delta}_t^{(N)}) - T_N(\ts_1^N(t),\Delta_t) \bigl| \biggl| \ts_1^N(t) \biggr] \biggr] \nonumber \\
&\leq \lim_{N\rightarrow\infty} E\biggl[ \sup_{F \in {\cal F}} \bigl|F(\tilde{\Delta}_t^{(N)}) - F(\Delta_t) \bigl| \biggr] \nonumber \\
&= 0. \nonumber
\end{align}
Hence, (\ref{neweq1}) implies (\ref{neweq2}) for any $t$.

Now, we prove that (\ref{neweq1}) is true for all $t$, which will complete the proof as (\ref{neweq1}) implies (\ref{neweq2}). Set $\sup_{N\geq1} \|g_N\| \eqqcolon L <\infty$ and define
\begin{align}
l_{N,t}(s,\Delta) \coloneqq \int_{\sA \times \sS_{t+1}} g_N(s') P_t(ds'|s,a,\Delta) \tpi_t^{(N)}(da|s). \nonumber
\end{align}
For any $s \in \sS_t$ and $(\Delta,\Delta') \in \P(\sS_t)^2$, we have
\begin{align}
&|l_{N,t}(s,\Delta) - l_{N,t}(s,\Delta')| \nonumber \\
&\phantom{xxxxxxxxxx}\leq L \sup_{(s,a) \in \sS_t \times \sA} \|P_t(\,\cdot\,|s,a,\Delta) - P_t(\,\cdot\,|s,a,\Delta')\|_{TV} \nonumber \\
&\phantom{xxxxxxxxxx}\leq L \omega_{P_t}(d_t(\Delta,\Delta')). \nonumber
\end{align}
Since $\omega_{P_t}(r) \rightarrow0$ as $r\rightarrow0$ by (III), the family $\{l_{N,t}(s,\,\cdot\,): s \in \sS_t, N\geq1\}$ is equicontinuous.

We prove (\ref{neweq1}) by induction on $t$. The claim trivially holds for $t=0$ as ${\cal L}(\ts_1^N(0)) = {\cal L}(\hs^N(0)) = \lambda_0$ for all $N\geq1$. Suppose the claim holds for $t$ and consider $t+1$. We can write
\begin{align}
&\bigl| {\cal L}(\ts_1^N(t+1))(g_N) - {\cal L}(\hs^N(t+1))(g_N) \bigr| \nonumber \\
&\phantom{xxxxx}=\biggl| \int_{\sS_t \times \P(\sS_t)} l_{N,t}(s,\Delta) {\cal L}(\ts_1^N(t),\tilde{\Delta}_t^{(N)})(ds,d\Delta) - \int_{\sS_t \times \P(\sS_t)} l_{N,t}(s,\Delta) {\cal L}(\hs^N(t),\delta_{\Delta_t})(ds,d\Delta) \biggr|. \nonumber
\end{align}
Since the family $\{l_{N,t}\}_{N\geq1}$ is equicontinuous and bounded, and (\ref{neweq1}) implies (\ref{neweq2}) at time $t$, the last term converges to zero as $N\rightarrow\infty$. This completes the proof.

\section*{Acknowledgments.}
This research was supported in part by the U.S.\ Air Force Office of Scientific Research (AFOSR) under MURI grant FA9550-10-1-0573, and in part by the Office of Naval Research under (ONR) MURI grant N00014-16-1-2710 and grant N00014-12-1-0998.


\end{document}